%% file: main.tex
\documentclass[sigconf]{acmart}

\AtBeginDocument{%
  \providecommand\BibTeX{{%
    \normalfont B\kern-0.5em{\scshape i\kern-0.25em b}\kern-0.8em\TeX}}}

\copyrightyear{2023} 
\acmYear{2023} 
\setcopyright{acmlicensed}\acmConference[CHI '23]{Proceedings of the 2023 CHI Conference on Human Factors in Computing Systems}{April 23--28, 2023}{Hamburg, Germany}
\acmBooktitle{Proceedings of the 2023 CHI Conference on Human Factors in Computing Systems (CHI '23), April 23--28, 2023, Hamburg, Germany}
\acmPrice{15.00}
\acmDOI{10.1145/3544548.3580852}
\acmISBN{978-1-4503-9421-5/23/04}
\input{macros}

\usepackage{xspace}
\usepackage[ruled,linesnumbered]{algorithm2e}    % for algorithm block
\usepackage{colortbl}
\usepackage{subcaption}
\usepackage{enumitem}

\makeatletter
\DeclareRobustCommand\onedot{\futurelet\@let@token\@onedot}
\def\@onedot{\ifx\@let@token.\else.\null\fi\xspace}

\def\eg{e.g\onedot}

\def\etal{et al\onedot}
\makeatother

\newcommand{\responseref}[0]{\color{black}}
\newcommand{\responseline}[1]{\textcolor{black}{#1}}
\newcommand{\camerareadyrevision}[1]{\textcolor{black}{#1}}
\begin{document}

%\title{\tool : Visualizing, Understanding, and Debugging Recurrent Neural Networks}
\title{\tool : Interactive RNN Explanation and Debugging via State Abstraction}

\author{Zhijie Wang}
\affiliation{%
  \institution{University of Alberta}
  \city{Edmonton}
  \state{AB}
  \country{Canada}
}
\email{zhijie.wang@ualberta.ca}

\author{Yuheng Huang}
\affiliation{%
  \institution{University of Alberta}
  \city{Edmonton}
  \state{AB}
  \country{Canada}}
\email{yuheng18@ualberta.ca}

\author{Da Song}
\affiliation{%
  \institution{University of Alberta}
  \city{Edmonton}
  \state{AB}
  \country{Canada}}
\email{dsong4@ualberta.ca}

\author{Lei Ma}
% \authornote{Lei Ma is also affiliated with Alberta Machine Intelligence Institute (Amii), Canada.}
\affiliation{%
  \institution{University of Alberta, Canada}
  \institution{The University of Tokyo, Japan}
  \city{}
  \state{}
  \country{}
  }
\email{ma.lei@acm.org}

\author{Tianyi Zhang}
\affiliation{%
  \institution{Purdue University}
  \city{West Lafayette}
  \state{IN}
  \country{USA}}
\email{tianyi@purdue.edu}

% \thanks{Zhijie Wang, Yuheng Huang, Da Song, and Lei Ma are also affiliated with the Alberta Machine Intelligence Institute (Amii), Canada.}
\begin{abstract}
Recurrent Neural Networks (RNNs) have been widely used in Natural Language Processing (NLP) tasks given its superior performance on processing sequential data. However, it is challenging to interpret and debug RNNs due to the inherent complexity and the lack of transparency of RNNs. While many explainable AI (XAI) techniques have been proposed for RNNs, most of them only support local explanations rather than global explanations. %Furthermore, it is not clear to what extend these explanations can help model developers. 
In this paper, we present \tool, an interactive system that provides both global and local explanations of RNN behavior in multiple tightly-coordinated views for model understanding and debugging. The core of {\tool} is a state abstraction method that bundles semantically similar hidden states in an RNN model and abstracts the model as a finite state machine. Users can explore the global model behavior by inspecting text patterns associated with each state and the transitions between states. Users can also dive into individual predictions by inspecting the state trace and intermediate prediction results of a given input. A between-subjects user study with 28 participants shows that, compared with a popular XAI technique, LIME, participants using {\tool} made a deeper and more comprehensive assessment of RNN model behavior, identified the root causes of incorrect predictions more accurately, and came up with more actionable plans to improve the model performance.
\end{abstract}




%   \centering
%   \includegraphics[width=\linewidth]{TeaserFigurePrint.pdf}
%   \caption{\textbf{\tool, an interactive system for visualizing, understanding, and debugging RNN models.} \textbf{(A) The \textit{State Diagram View}} displays the abstracted states and transitions of an RNN model. \textbf{(B) The \textit{Pattern Summary View}} displays common text patterns learned by an RNN model. \textbf{(C) The \textit{Instance View}} displays the raw data as an interactive data grid for user to explore the raw data used to train or test an RNN model. \textbf{(D) \textit{Intermediate Prediction Results}} are rendered when users input a new sentence.}
%   \Description{The interface of \tool}

\begin{teaserfigure}
  \centering
  \includegraphics[width=0.75\linewidth]{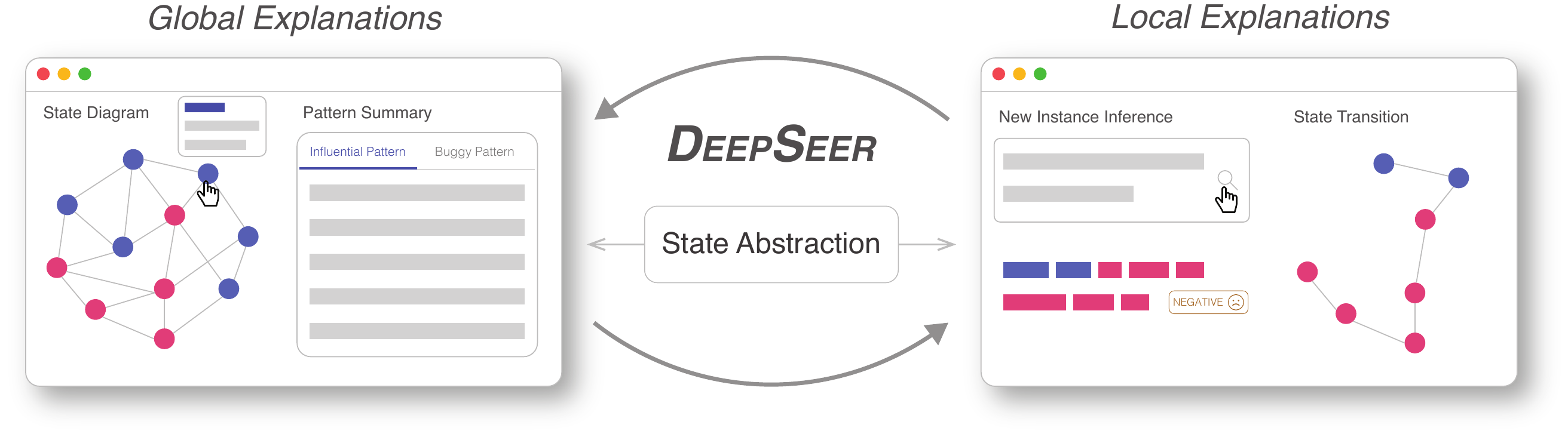}
  \caption{\textbf{{\tool} is an interactive tool for supporting RNN model understanding and debugging via state abstraction.} {\tool} helps programmers by providing \textit{Global Explanations} and \textit{Local Explanations} as a synergistic loop for an RNN model. Programmers can use {\tool} to quickly understand and identify potential bugs by exploring \textit{Global Explanations}, then zoom into \textit{Local Explanations} to contextualize global explanations. Programmers can also debug on a specific instance according to \textit{Local Explanations}, then validate their debugging hypothesizes by zooming out to compare with \textit{Global Explanations}.}
  \Description{The interface of \tool}
  \label{fig:teaser_ui}
  \Description{This figure shows the basic idea of DeepSeer, which contains two subfigures. The left one shows how DeepSeer renders the model’s global explanation by displaying the state diagram and pattern summaries. The right one shows how DeepSeer renders the model’s local explanation by providing intermediate prediction results and state traces. These two kinds of explanations are unified with the state abstraction technique.}
\end{teaserfigure}

\sloppy 
\maketitle

\textcolor[HTML]{d24d00}{WARNING: The toxicity detection example in the usage scenario contains some content that might be distracting.}

\input{intro}

\input{background}

\input{related}

\input{design}

\input{approach}

\input{scenario}

\input{user-study}

\input{results}

\input{discussion}

\input{conclusion}

\begin{acks}
We would like to thank all anonymous participants in the user study and anonymous reviewers for their valuable feedback. This work was supported in part by Amii RAP program, Canada CIFAR AI Chairs Program, the Natural Sciences and Engineering Research Council of Canada (NSERC No.RGPIN-2021-02549, No.RGPAS-2021-00034, No.DGECR-2021-00019), as well as JSPS KAKENHI Grant No.JP20H04168, JST-Mirai Program Grant No.JPMJMI20B8.
\end{acks}

\balance

% \section{Acknowledgments}

% Identification of funding sources and other support, and thanks to
% individuals and groups that assisted in the research and the
% preparation of the work should be included in an acknowledgment
% section, which is placed just before the reference section in your
% document.

% \section{Appendices}

% If your work needs an appendix, add it before the
% ``\verb|\end{document}|'' command at the conclusion of your source
% document.

% Start the appendix with the ``\verb|appendix|'' command:
% \begin{verbatim}
%   \appendix
% \end{verbatim}
% and note that in the appendix, sections are lettered, not
% numbered. This document has two appendices, demonstrating the section
% and subsection identification method.

\bibliographystyle{ACM-Reference-Format}
\bibliography{reference}

\input{appendix}

% \appendix









\end{document}

%% file: macros.tex
\newcommand{\tool}{\textsc{DeepSeer}}

\newcommand{\circled}[1]{{\large \textcircled{\footnotesize #1}}}

\usepackage{xcolor}

\definecolor{primary}{HTML}{3f51b5}
\definecolor{secondary}{HTML}{f44336}

\usepackage{multirow}

%% file: intro.tex
\section{Introduction}
\label{sec:intro}

Deep neural networks (DNNs) have been increasingly adopted in practice due to their superior performance on real-world challenging tasks, e.g., self-driving~\cite{peng2020first}, virtual assistant~\cite{bocklisch2017rasa}, and disease diagnosis~\cite{li2014deep}. The rapid development of deep learning systems brings opportunities, but also challenges and concerns. One of the major concerns arises from the interpretability of DNNs~\cite{lipton2018mythos}. Unlike traditional software whose decision logic is manually programmed in the form of source code, a DNN model includes a large number of neurons connected by non-linear functions, whose weights are automatically learned from training data. The internal states of traditional software can be easily inspected and analyzed by setting breakpoints and checking runtime values. However, the internal states of a DNN model are high-dimensional vectors rather than symbolic values. It is hard to tell what kinds of patterns a DNN model has learned by inspecting these vectors or why the model makes a specific prediction. Therefore, this internal complexity and inscrutability of DNNs lead to significant debugging challenges, as well as concerns about the trustworthiness and reliability of DNNs.

Although there is a recently growing interest in improving the interpretability of DNNs in the ML, HCI, and Visualization communities, many existing techniques treat a DNN model as a black box and generate model-agnostic explanations such as feature importance, without revealing the inner workings of the DNN model~\cite{ribeiro2016should, lundberg2017unified,koh2017understanding, ribeiro2018anchors}. While there are some techniques for visualizing the hidden states in a DNN model, many of them focus on convolutional neural networks (CNNs)~\cite{bau2017network, simonyan2013deep, olah2018building}. In this work, we are particularly interested in recurrent neural networks (RNNs). Compared with other kinds of DNNs, recurrent neural networks (RNNs) are capable of processing sequential data with variable lengths, such as text and audio. The recurrent architecture affords an internal memory in RNNs, which is proven effective for learning temporal patterns in sequential data. Yet this architecture also poses challenges in visualizing the internal states of RNNs. Unlike CNNs which have a fixed number of layers and neurons in each layer, RNNs are {\em unbounded}. Furthermore, instead of treating each layer separately, which is a common practice in CNN visualization, it is important to visualize the dynamics of RNN units, i.e., the temporal patterns embedded in a sequence of hidden states. 

\responseline{In this paper, we present {\tool}, an interactive system that allows model developers to understand and debug RNN models. Our key insight is to treat an RNN model as a stateful system. By clustering and abstracting semantic similar hidden states, an RNN model can be represented as a finite-state machine (FSM), which is much smaller and more navigable compared with the original RNN model. Furthermore, instead of directly visualizing the values of hidden states as in prior work~\cite{strobelt2017lstmvis}, {\tool} projects hidden states to a more interpretable representation---the common words and phrases associated with these states. By inspecting the transitions among states, users can quickly identify the temporal patterns learned by the model.} 

\responseline{To assess the overall usefulness of {\tool}, we conducted a between-subjects user study with 28 programmers of various levels of expertise in ML and RNNs. Given a pre-trained RNN model, participants were asked to complete a model understanding task followed by a debugging task using either {\tool} or a popular XAI tool, LIME~\cite{ribeiro2016should}. We found that in the model understanding task, participants using {\tool} provided more insightful answers about the model behavior, pinpointed model limitations more precisely, and gave more useful and diverse suggestions about how to improve the assigned model. Furthermore, in the model debugging task, participants using {\tool} identified the reasons for the misclassifications more correctly than participants using LIME.} 

In summary, this work makes the following contributions:

\begin{itemize}[leftmargin=*]
    \item {\bf System.} We design and develop an interactive system for understanding and debugging the internal behavior of RNNs. We first leverage the state abstraction method to abstract an RNN model as a finite state machine through bundling semantically similar hidden states. Then we design and implement three tightly-coordinated views: state diagram view, pattern summary view, and instance view to visualize and interpret the internal behavior of an RNN model from different perspectives. We have open-sourced our system on GitHub~\footnote{\href{https://github.com/momentum-lab-workspace/DeepSeer}{https://github.com/momentum-lab-workspace/DeepSeer}}.
    
    \item {\bf Visualizations and interactions.} We propose a set of visualization and interaction designs to facilitate the interpretation and debugging of RNNs at different granularities. Specifically, we combined state diagrams, responsive tooltips, state traces, color highlighting, filtering, instance matching, and pattern summarization to simultaneously show the global model behavior, instance-level explanations, critical patterns, and similar instances. 
    
    \item {\bf Evaluation.} A between-subjects user study demonstrates the usefulness of {\tool} to ML developers when understanding the overall behavior of a model and debugging misclassifications.
\end{itemize}

%% file: background.tex
% \iffalse
\section{Background: Recurrent Neural Networks}
\label{sec: background}

Recurrent Neural Networks (RNNs) are a type of deep neural network that is specifically designed for processing sequential input, e.g., text data. In this section, we briefly introduce the basics of it.

\begin{figure}[t]
    \centering
    \includegraphics[width=0.8\linewidth]{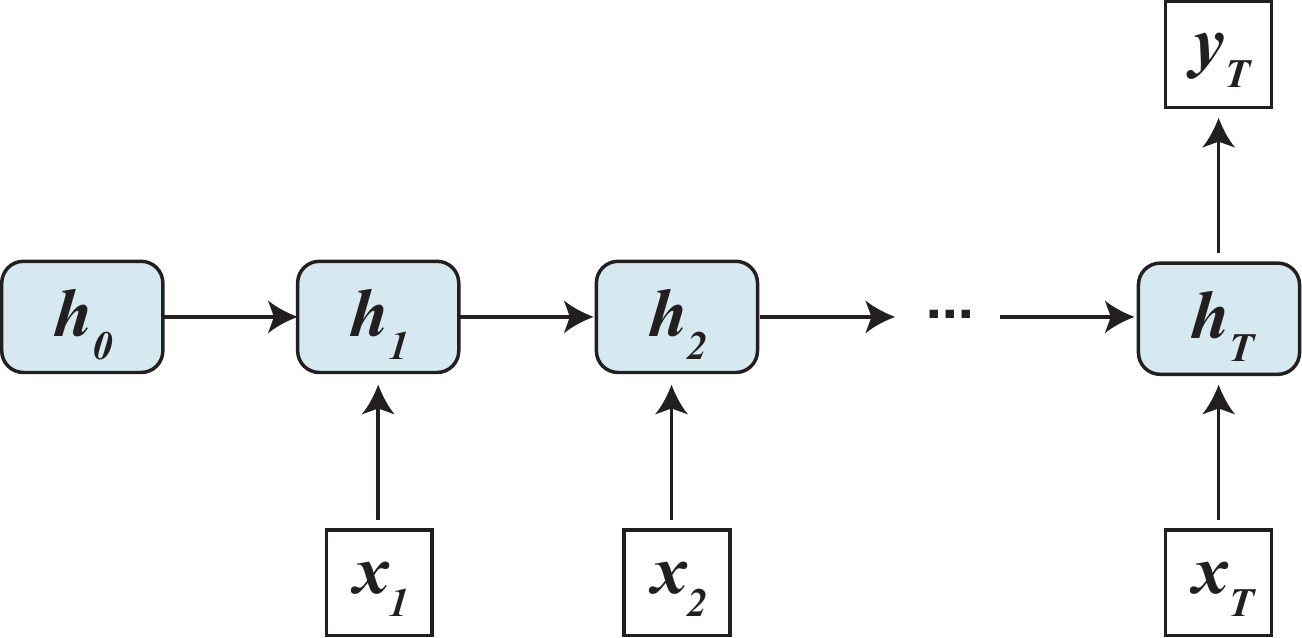}
    \caption{\responseref{}The workflow of a basic recurrent neural network (RNN). At each time step $t$, RNN takes an input $x_t$ to update its hidden state $h_t$. The prediction result at time step $t$ ($y_t$) is obtained by processing the hidden state $h_t$.}
    \Description{This figure shows the workflow of a basic recurrent neural network (RNN). At the beginning, an RNN model first initializes a hidden state vector h0. Then at each time step t, the RNN model takes one input x_t, and updates the hidden state vector as h_t. The output of this RNN model is obtained by transforming the last hidden state vector into y.}
    \label{fig:RNN}
\end{figure}

\sloppy As shown in Fig.~\ref{fig:RNN}, an RNN model takes sequential inputs $\{x_1, x_2, \dots, x_T\}$. The RNN model first initializes its hidden state vector $h_0\in \mathbb{R}^N$, where $N$ is the dimension of this hidden state vector. At a time step $t$, the RNN model takes an input $x_t (1\leq t\leq T)$ to update its internal hidden state from the last time step $h_{t-1}$ to the new hidden state $h_t$. This process can also be seen as maintaining and updating the ``hidden memory'' of an RNN model. Therefore, to understand an RNN model, it is important to interpret such ``memory'' maintained in different hidden states~\cite{strobelt2017lstmvis, ming2017understanding}.

To make a prediction at time step $t$, an RNN model transforms the hidden state $h_t$ into the desired output format $y_t$. For instance, to perform a classification task, the hidden state $h_t$ is usually fed into an MLP (multilayer perceptron) network to project $h_t$ into $u_t\in\mathbb{R}^K$, where $K$ is the number of classes. Then a probability distribution $p_t$ is computed through a ``softmax'' function:

\begin{equation}
    \label{eq:softmax}
    \begin{aligned}
        p_t &= \text{softmax}(u_t) \\
        p_t^i &= \frac{e^{u_t^i}}{\sum_{j=1}^K e^{u_t^j}}\quad\text{for }i=1,\dots,K
    \end{aligned}
\end{equation}
The prediction result at time step $t$ is further computed by finding a label $k$ which produces the maximum probability $p_t^k$.

Note that the process of updating the hidden state $h_t$ can be achieved by different types of RNN units, such as the Elman RNN cell~\cite{elman1990finding}, long short-term memory (LSTM)~\cite{hochreiter1997long}, and gated recurrent unit (GRU)~\cite{cho2014properties}. In our user study sessions, we use GRU, which shows better efficiency compared with other variants. Note that our proposed system only requires access to the hidden states and does not require access to the updating process inside an RNN unit. Therefore, it can be generalized to different types of RNN units.

% \fi

%% file: related.tex
\section{Related Work}
\label{sec:related_work}

\subsection{Explainable AI}
%Zhijie: I think it might be better to organize this paragraph by following some kind of concepts in our design.

Our work is most related to Explainable AI (XAI), since it promotes model interpretability by abstracting a DNN model as a finite state machine (i.e., a global explanation) and by rendering the state trace of a given instance (i.e., a local explanation). Previous studies have shown that supporting model interpretability can increase user acceptance and trust of the system~\cite{herlocker2000explaining, dzindolet2003role, schneider2021explain, kizilcec2016much}, improve fairness~\cite{dodge2019explaining}, and improve human-AI team performance~\cite{cai2019effects}. %Although there is no formal or agreed definition of interpretability~\cite{lipton2018mythos, hohman2019gamut}, previous discussions have made a distinction between interpretation (\textit{interpretability}) and explanation (\textit{explainability})~\cite{montavon2018methods}. 
A good interpretation should be in an interpretable domain~\cite{montavon2018methods}, i.e., mapping any of abstract concepts (e.g., numeric vectors) into a domain (e.g., images, texts) that the human can understand. Our work is inspired by this principle---instead of visualizing hidden state values as in some prior work~\cite{strobelt2017lstmvis}, we map hidden states back to linguistic patterns in the text corpus. In this way, users can easily recognize what patterns an RNN model has learned from the training data. %Upon the interpretation, an explanation should answer ``why does this specific input lead to that specific output?''~\cite{gilpin2018explaining}
%XAI has drawn much attention these years, especially after the release of the European General Data Protection Regulation (GDPR)~\cite{goodman2017european}. Fan et.al. ~\cite{fan2021interpretability} show that there is an exponential growth in the number of articles in the field. The reasons can be many. From the usability perspective, prior studies have shown that providing explanations for complex autonomous systems can increase the acceptance of the systems~\cite{ , dzindolet2003role}. There are also some discussions on the trust~\cite{lui2018artificial, hengstler2016applied}, bias~\cite{chen2019hidden, sinz2019engineering}, fairness~\cite{hoffmann2019fairness, dodge2019explaining}, and ethics~\cite{cath2018artificial, etzioni2017incorporating} issues of AI, indicating the need for more transparency. Furthermore, supporting interpretability can also help developers understand model behavior and improve model performance~\cite{hohman2019gamut}.

% hastie2017generalized -- mode-agnostic delete

% \todo{add citations for model-agnostic method}
Existing XAI methods can be roughly grouped into two categories: model-agnostic methods and model-aware methods.
%, attribution-based methods, influence function methods, and example-based methods. 
Model-agnostic methods~\cite{ribeiro2016should, lundberg2017unified, ribeiro2018anchors} treat the model to be explained as a black box. LIME~\cite{ribeiro2016should} is a well-known technique in this category. Given an input instance, it learns a simpler and interpretable model (also known as a surrogate model), such as a linear regression model, to approximate a complex model using the training data near the given instance.  By rendering the feature's importance in the surrogate model, LIME generates a local explanation for the prediction of the given instance. However, these model-agnostic methods usually ignore the internal behavior 
of a model when generating explanations. Specifically, given an RNN model, they do not take the transitions between different hidden states into account. On one side, this may lead to low-fidelity explanations~\cite{rahnama2019study}. On the other hand, advanced user groups such as model developers may find it insufficient to debug model behavior~\cite{strobelt2017lstmvis}. To address this challenge, {\tool} is designed to investigate the internal behavior of an RNN model via a novel finite state machine abstraction.
%To address this issue, {\tool} generates temporal patterns as explanations, i.e., \textit{influential patterns} and \textit{possible buggy patterns}, by leveraging the state transitions behind an RNN model.
%SHAP is built upon LIME and has some mathematical properties %(Local accuracy, Missingness, Consistency) that LIME doesn't have. 
%GAMs ~\cite{hastie2017generalized} as a model-agnostic method can provide global explanations. It can be thought of as a generalized linear model with the ability to model nonlinearity between features and output. GAMUT~\cite{hohman2019gamut} has successfully applied it to offer global explanations. 

Unlike model-agnostic methods, model-aware methods try to open up the {\em black box} of a DNN. Among different model-aware methods, our work is most related to attribution-based methods and influence function methods. Attribution-based methods~\cite{zeiler2014visualizing, smilkov2017smoothgrad, selvaraju2017grad} often use activation or gradient information in a DNN model to compute the importance score for input features, e.g., pixels in an image, tokens in a sentence. For example, Karpathy et.al.~\cite{karpathy2015visualizing} presents a visualization that maps neurons' activation to individual characters in a sentence. This visualization is only applicable to individual sentences (i.e., local explanations), which becomes hard to interpret with a large number of sentences. Our work differs in a way that we aggregate words and phrases with similar hidden states from many sentences in a finite state diagram while also providing a way to delve into the state trace of individual sentences.
Influence function methods~\cite{koh2017understanding, barshan2020relatif, KohATL19} compute the influence of an individual training instance based on gradients and identify a set of instances that have a big impact on model predictions. %In contrast to the above methods trying to compute the score, feature-based methods~\cite{olah2017feature, olah2018building, bau2017network, karpathy2015visualizing} use the activation maximization technique to find abstract features learned from the model. These features can be visualized for humans to understand the model better. 
Our design of {\em influential patterns} and \textit{possible buggy patterns} draws inspirations from these methods. Specifically, {\tool} summarizes short text patterns which usually significantly affect model predictions or have led to possible bugs by analyzing the hidden states of training data. %In addition, {\tool} reveals the influence of training instances through interpreting on the hidden states instead of gradient information. 
%Moreover, the audience of {\tool}--model developers--might use these global, aggregate explanations for debugging and further improvement~\cite{hohman2019gamut}. Therefore we argue that only summarizing a limited number of isolated representative examples are not enough for model developers. Hence, once the user of {\tool} identifies some of representative examples (through \textbf{Pattern Summary View}), they can also easily find similar training instances in the \textbf{Instance View}. By taking the state trace as a ``bridge'', users of {\tool} don't only find instances containing the exact same patterns, but also patterns with similar semantic meanings.

% There are also some other kinds of XAI methods, e.g., representative examples~\cite{adhikari2019leafage, kim2016examples}, counterfactuals~\cite{byrne2019counterfactuals, wexler2019if, wu2021polyjuice}, adversarial examples~\cite{goodfellow2014explaining, su2019one}, etc. Please refer to existing surveys and literature reviews~\cite{molnar2020interpretable, arrieta2020explainable, adadi2018peeking} for more details.

\responseline{We further refer readers to existing surveys and literature reviews~\cite{molnar2020interpretable, arrieta2020explainable, adadi2018peeking} for more details about different XAI methods.
}

\subsection{DNN Debugging, Testing, and Repairing}
Several explainable AI techniques have been used to understand and debug model errors~\cite{ribeiro2016should, kim2018interpretability, koh2017understanding, adebayo2020debugging}. For example, Ribeiro et.al.~conducted a user study with  27 participants and showed that the explanations generated by LIME could be used to detect spurious correlations learned by a model. %Kim et.al.~\cite{kim2018interpretability} introduce Concept Activation Vectors (CAVs) to interpret DNNs and use the explanation for diagnosing spurious correlation. 
Koh et.al.~\cite{koh2017understanding} have shown that influence functions can be used to debug domain mismatch. However, Adebayo et.al.~\cite{adebayo2020debugging} found that post-hoc model explanations, especially attribution-based methods, are sometimes ineffective for detecting certain kinds of bugs such as label error and out-of-distribution error.

In parallel, the Software Engineering (SE) community has developed several techniques by adapting traditional SE techniques to debug, test, and repair DNN models~\cite{yu2020deeprepair, xie2021rnnrepair, ma2018mode, ma2017lamp}. %To get an overview of the field we refer the reader to the detailed survey~\cite{zhang2020machine}. 
DeepRepair~\cite{yu2020deeprepair} uses a style-transfer-based data augmentation method to repair DNN models. RNNRepair~\cite{xie2021rnnrepair} identifies influential instances for retraining and remediates two types of incorrect predictions at the sample and segment levels.  MODE~\cite{ma2018mode} presents a debugging workflow by first conducting model state differential analysis and then selecting training instances for retraining. LAMP~\cite{ma2017lamp} provides data provenance information by computing the importance of input through automated differentiation. These techniques focus on automating the debugging and retraining pipeline and do not involve humans in the loop. %Our work differentiates from them as our method is human-in-the-loop, and the traceability of models we provide for users can be beneficial for both model debugging and data debugging, whereas it is harder for them to find data bug. 
Our work differs from these techniques in two ways. First, {\tool} aims to provide a comprehensive understanding of model behavior by abstracting RNNs as a state diagram and identifying influential patterns, going beyond diagnosing model prediction errors. Second, to enable model developers to diagnose model errors, {\tool} renders the intermediate prediction results of input and provides affordances for investigating how individual words and phrases influence the prediction result, rather than automatically localizing the root cause of a model error.

\subsection{RNN Visualization}
Many DNN visualization techniques have been proposed to help users understand and analyze the inner workings of DNN models.
The most related visualization techniques to us are those specifically designed for RNNs~\cite{strobelt2017lstmvis, ming2017understanding, karpathy2015visualizing}. Karpathy et.al.~\cite{karpathy2015visualizing} visualizes which characters in an input sentence the RNN attends to based on the activation function output. Li et.al.~\cite{li2015visualizing} use a gradient-based salience score, rather than neuron activation, to measure the importance of each word in an input sentence. The salience scores are then visualized in a heatmap. Both visualizations are static and can only visualize the hidden states of a single input at a time. Strobelt et.al.~\cite{strobelt2017lstmvis} extend them by building an interactive visualization approach called LSTMVis. LSTMVis renders individual hidden states in a parallel coordinates plot. It allows users to interactively select specific segments of an input sentence and search for other inputs with similar hidden states. However, given that RNNs typically have hundreds or even thousands of hidden states, visualizing individual hidden states can lead to significant cognitive overhead for users. To address this issue, {\tool} clusters similar hidden states to an abstract state and represents an RNN model as a finite state machine. This significantly reduces the number of states users need to keep track of and also allows {\tool} to directly visualize the finite state machine to provide a global view of the entire model rather than individual hidden states. Our work is 
also related to RNNVis~\cite{ming2017understanding}. RNNVis clusters similar hidden states as memory chips and renders text inputs associated with each cluster as word clouds. However, unlike a finite state machine, this design does not capture the transition between hidden states or render longer linguistic patterns beyond common words. Furthermore, {\tool} provides additional features to facilitate model inspection and debugging, e.g., rendering intermediate prediction results, summarizing influential patterns and buggy patterns, etc. %Taking abstract states as a bridge, we provide users with both text-level and state-level visualizations. In addition, we allow users to verify their hypothesis on the model’s behaviors quickly by generating state diagram for user-defined sentences on the fly.

% Tianyi: the papers below are all from the visualization community. They are not specific for DNN models but mostly for mode comparision in general (agnostic to models). So we don't have to mention them.
%Manifold~\cite{zhang2018manifold} provides an interactive interface for interpreting, debugging, and comparing ML models by solely observing the input and the output. ModelTracker~\cite{amershi2015modeltracker} and Squares~\cite{ren2016squares} present interactive visualization for ML models' performance analysis and debugging. They both provide statistics summary and instance-level performance visualization. 

%% file: design.tex
{\responseref{}

\begin{figure*}[t]
  \centering
  \includegraphics[width=0.95\linewidth]{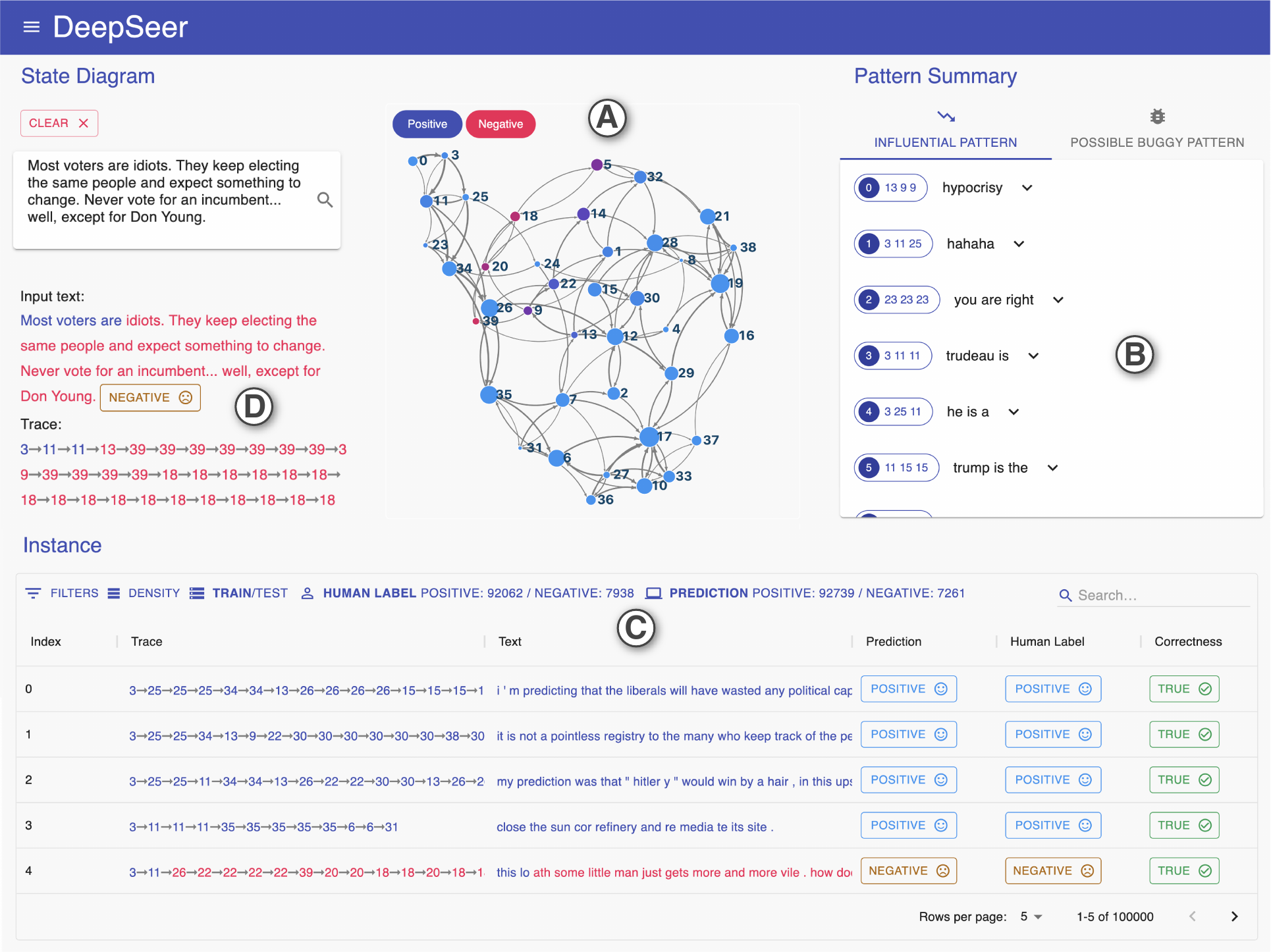}
  \caption{\textbf{\tool, an interactive system for visualizing, understanding, and debugging RNN models.} \textbf{(A) The \textit{State Diagram View}} displays the abstracted states and transitions of an RNN model. \textbf{(B) The \textit{Pattern Summary View}} displays common text patterns learned by an RNN model. \textbf{(C) The \textit{Instance View}} displays the raw data as an interactive data grid for users to explore data used to train or test an RNN model. \textbf{(D) \textit{Intermediate Prediction Results}} are rendered when users input a new sentence.}
  \label{fig:teaser}
  \Description{This figure shows the interface of DeepSeer, which includes three different views. (A) The first view is the State Diagram View, which locates in the top-left of the interface. This view displays the abstracted states and transitions of an RNN model as a directed graph. (B) The second view is the Pattern Summary View, which locates in the top-right of the interface. This view displays common text patterns learned by an RNN model as a two-column list. The left column displays the influential patterns, where each row is a specific influential pattern. The right column displays the buggy pattern. (C) The third view is the Instance View, which locates at the bottom of the interface. This view displays the raw data as an interactive data grid for users to explore the raw data used to train or test an RNN model. In this data grid, users can scroll to explore different data instances. Users can also filter the data according to specific conditions. }
\end{figure*}

\section{Design Goals and System Overview}
\label{sec:design}

In this section, we first summarize the design goals of our system based on a literature review. Then, we present a system overview to discuss how our system design supports each design goal.

\subsection{User Needs and Design Goals}

To understand the needs of RNN developers, we perform a literature review of previous papers that have done a formative study of interpreting DL models, have done a user study of existing tools, or have discussed the challenges and opportunities of explaining and debugging DL models. Based on the literature review, we summarize the following design goals for {\tool}: 

% \begin{itemize}
    \vspace{1mm}
    \noindent {\bf G1. Help users understand the overall behavior of an RNN model.} Previous studies have shown that model developers prefer to have a high-level understanding of what has been learned by the model~\cite{ming2017understanding, liao2020questioning, das2020opportunities}. 
    For instance, Kaur~\etal surveyed 197 ML developers about the interpretability tool's capabilities, and 61\% of responses mentioned the importance of global explanations~\cite{kaur2020interpreting}.
    Specifically, Ming~\etal highlighted the importance of rendering the semantic information captured by the hidden states of an RNN model~\cite{ming2017understanding}. 
    Thus, {\tool} should help model developers understand the overall behavior of an RNN model, especially the semantic information learned by its hidden states.  
    
    \vspace{1mm}
    
    \noindent {\bf G2. Help users understand the model decision-making process on a specific input of interest.} When inspecting individual prediction results, especially incorrect ones, model developers wish to understand why the model makes such a prediction on the particular input~\cite{liao2020questioning, amershi2019guidelines, hohman2019gamut}. For instance, through a formative study with nine ML developers,  Hohman~\etal~\cite{hohman2019gamut} found that users wanted to see how different features contributed to the model's decision. 
    Furthermore, Kahng~\etal interviewed fifteen Facebook developers and found that a natural way for them to understand complex models was to inspect the model behavior on individual examples~\cite{kahng2017cti}. The importance of local explanation is also confirmed by the large-scale survey~\cite{kaur2020interpreting}---65\% of respondents considered local explanations important. 
    
    \vspace{1mm}
    
    \noindent {\bf G3. Help and assist users in searching for similar data.} \sloppy Through a user-centered design process with two NLP developers, Liu et al.~found that NLP developers typically follow an ``exploration-centric'' approach to discover and debug errors in an NLP model~\cite{liu2018nlize}. That is, developers prefer to inspect and compare predictions among similar input examples to get insights. Therefore, traceability should also be provided to help users easily explore similar examples when debugging a model prediction~\cite{hohman2019gamut}. Specifically, Strobelt~\etal highlighted that matching similar examples for RNN could help developers validate an interpretation hypothesis~\cite{strobelt2017lstmvis}.
    
    \vspace{1mm}

    \noindent {\bf G4. Help users summarize the common characteristics of input data.} 
    Inspecting individual data points can be tedious and time-consuming, hindering insight discovery. Kahng et al.~found that model developers at Facebook often curated subsets of data with common characteristics to understand how a model behaves at high-level categorization~\cite{kahng2017cti}. Furthermore, helping users identify common input characteristics is particularly useful for error analysis. Jin~\etal found that ML developers usually needed to examine the characteristics shared by a set of wrong predictions and verify whether error patterns formed by these characteristics make sense~\cite{jin2022gnnlens}. However, this is often manually done by users based on their domain knowledge. Therefore, {\tool} should support users in identifying and examining common characteristics of inputs, especially mispredicted inputs.
    
% \end{itemize}

\subsection{System Overview}

To support users gaining a high-level understanding of what has been learned by an RNN model (\textbf{G1}), we choose to render an RNN model as a state diagram in which each node is a group of similar hidden states from the RNN model, as shown Figure~\ref{fig:teaser}~\circled{A}. Compared with the original RNN model, which has hundreds or thousands of hidden states, the state diagram is much smaller after state clustering and thus more navigable. Furthermore, {\tool} binds each node with the text patterns memorized by the corresponding hidden states to help users interpret the semantic meaning of the hidden states. Compared with an alternative design of directly visualizing the hidden states values~\cite{strobelt2017lstmvis, karpathy2015visualizing}, which are high-dimensional arrays and hard to interpret, the state diagram is easier to navigate and inspect.

To help users understand the model decision-making process on specific inputs (\textbf{G2}), {\tool} visualizes the intermediate model prediction result after an RNN model reads each word in an input sentence (Figure~\ref{fig:teaser}~\circled{D}). In this way, users can easily see which word sways the decision of the model and contributes more to the final result. To support \textbf{G3}, {\tool} allows users to search input sentences with {\em similar text patterns} (i.e., have the same keyword or follow the same regular expression) or with {\em similar model behavior pattern} (i.e., have the same state or follow the same state trace) in an instance view (Figure~\ref{fig:teaser}~\circled{C}). To help users find common patterns (\textbf{G4}), {\tool} proactively identifies frequent text patterns that have a high influence on model prediction results, as well as patterns that are shared among incorrect predictions (Fig.~\ref{fig:teaser}~\circled{B}). Such common patterns can also serve as a complementary global explanation method (\textbf{G1}), since it provides more straightforward starting points for investigation if users find a state diagram overwhelming. 

}

%% file: approach.tex
\section{Design and Implementation}
\label{sec:approach}

\subsection{State Abstraction}
\label{subsec:abstraction}

\begin{figure*}[t]
\centering
  \includegraphics[width=0.85\linewidth]{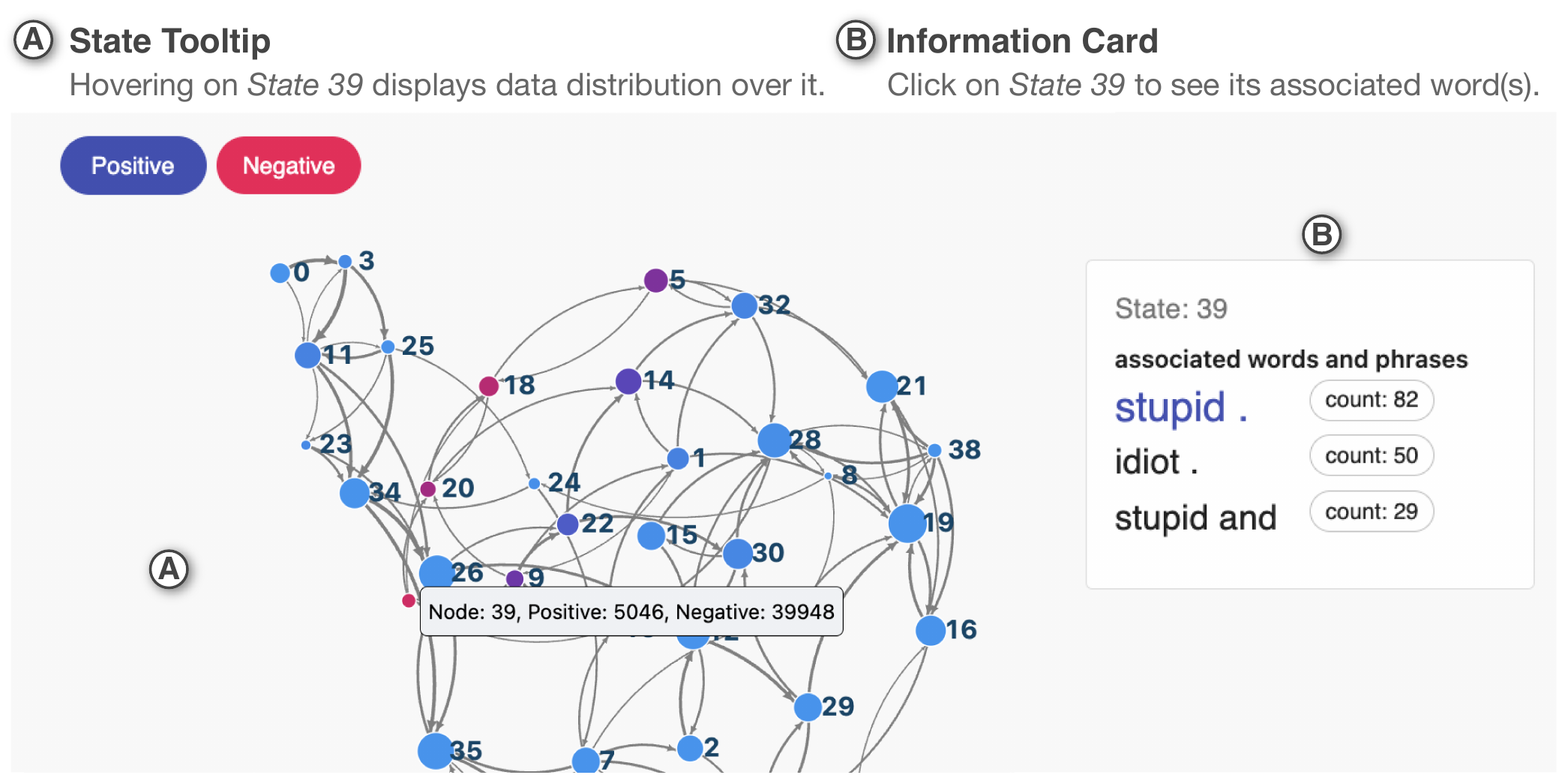}
  \caption{Interacting with the \textbf{\textit{State Diagram View}} to understand the abstracted states of an RNN model.} 
  \Description{This figure displays the state diagram view and two kinds of interactions with it. The first is when users hover over a specific state, i.e., state 20 in the figure, they can find that in the training data, there are 25342 times this state is classified as positive, while 40319 times as negative. The second interaction is when users click on a specific state, i.e., state 20 in the figure, they can inspect an information card on the right of the state diagram view, which shows the associated words and phrases of this state.}
  \label{fig:state_interact}
\end{figure*}

To generate a state diagram from an RNN model, we develop a method that clusters semantically related hidden states of the RNN model into an abstract state. Our work is inspired by the model-based analysis of stateful RNNs~\cite{du2019deepstellar, du2020marble, weiss2018extracting, okudono2020weighted, rabusseau2019connecting}. These works apply various techniques to extract interpretable state transition models (e.g., discrete-time Markov chain, automata) from stateful RNNs. By turning complex RNNs into interpretable state transition models, black boxes are turned into more transparent models and thus improve the model interpretability, which also provides the possibility for further analysis. 
We choose to build on top of a state-of-the-art method, DeepStellar~\cite{du2019deepstellar}, since it is demonstrated to be effective in various tasks, including adversarial detection~\cite{du2019deepstellar}, DNN testing~\cite{du2019deepstellar}, and DNN repair~\cite{xie2021rnnrepair}. Previous work has also shown that abstracted states can make the same prediction as the original RNN model in 97\% of test data~\cite{xie2021rnnrepair}.

To obtain the FSM representation for a trained RNN, we abstract over both the states and the transitions. \responseline{Appendix~\ref{appendix:state_abstraction} presents the algorithm for state abstraction. Here we briefly summarize how it works. For each instance in the training data, our method first records the intermediate hidden vectors $\{h_1, h_2, \dots, h_l\}$ during inference, where $h_i$ ($1\leq i \leq l$) is a concrete hidden state of an RNN model. $l$ denotes the number of tokens in a sentence.} Our method then applies Principle Component Analysis (PCA) for dimension reduction on all concrete states collected from training data before abstraction. Different from DeepStellar~\cite{du2019deepstellar}, which uses an interval-based method for states abstraction, our method applies Gaussian Mixture Model (GMM)~\cite{mclachlan1988mixture} to cluster similar concrete states. GMM addresses two key limitations in the interval-based method: 
1) newly generated hidden vectors can fall outside the interval at test time, resulting in unknown states; 2) the number of states grows exponentially with $k$ dimension and $m$ intervals ($m^k$), and too many states can be hard to visualize.
With state abstraction, the model prediction process on a given input can be modeled as a sequence of abstract states. We call this state sequence the {\em trace} of model prediction.

\responseline{We conducted a quantitative analysis of the faithfulness of state abstraction. We measured the prediction consistency between the abstracted and original models in the three different NLP tasks from the usage scenario (Section~\ref{sec:scenario}) and the user study (Section~\ref{subsec:tasks}). The prediction consistency on the test data is 99\%, 97\%, and 85\%, respectively. This implies that abstracted models can faithfully represent the behavior of RNNs. Appendix~\ref{appendix:faithfulness} includes the experiment details. }

\responseline{Different from the previous work focusing on state abstraction technique itself or using the technique for model testing and repairing, our work is the first to extend this technique for interactive model explanation and debugging with a more accessible user interface. Our work integrates state abstraction into a ``human-in-the-loop'' approach for the first time to support users in understanding and debugging an RNN model with rich interaction mechanisms. In the following subsections, we will introduce the interactive features of {\tool} built on top of state abstraction.}

\begin{figure*}[t]
  \includegraphics[width=0.8\linewidth]{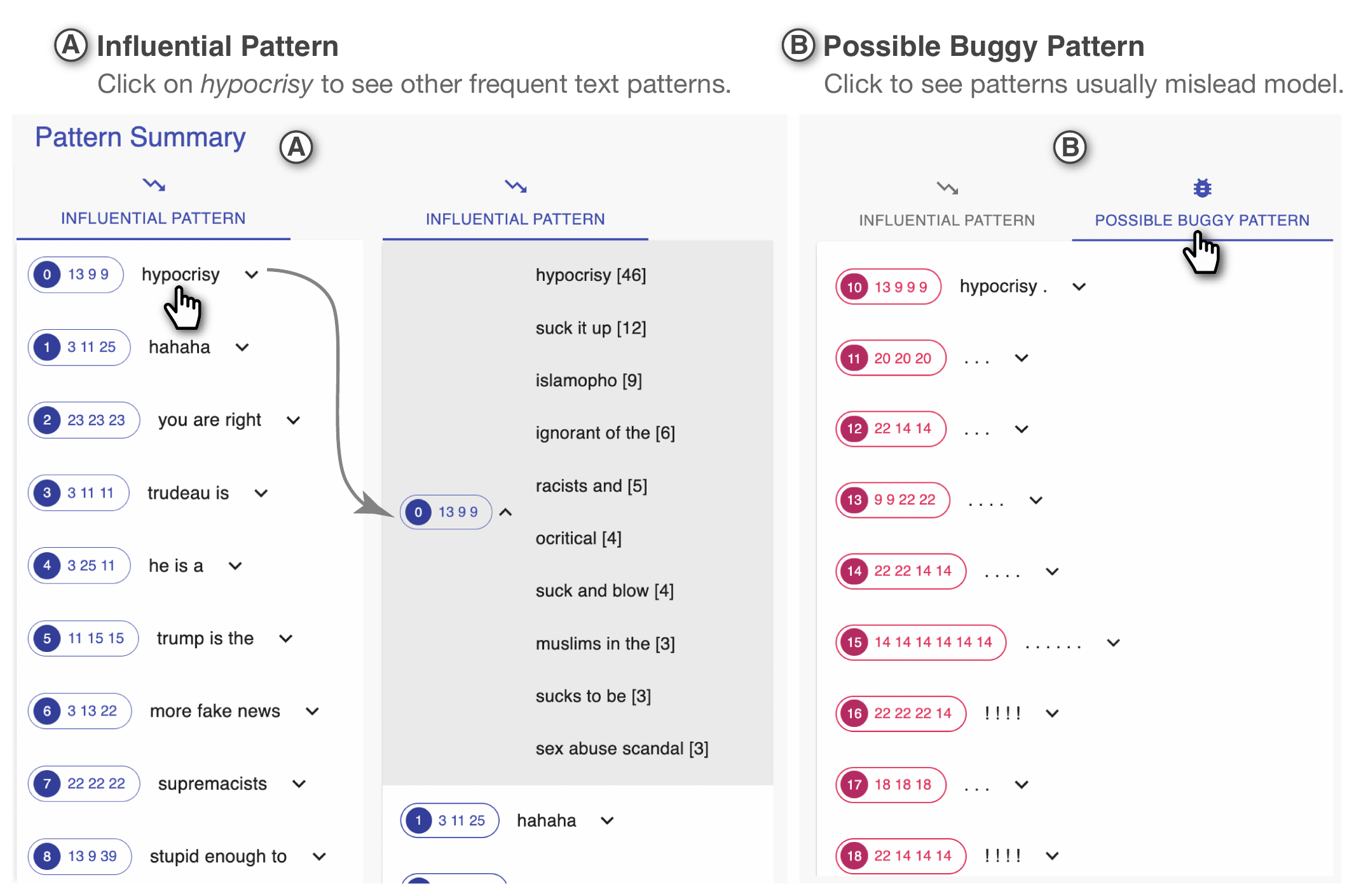}
  \caption{Interacting with the \textbf{\textit{Pattern Summary View}} in \tool{} to inspect the text patterns learned by an RNN model.} 
  \Description{This figure has two subfigures. The first one shows the influential patterns. The influential patterns identified by this RNN model include “hypocrisy”, “hahaha”, “you are right”, etc. When users click on a specific pattern, i.e., hypocrisy, they can see more influential patterns sharing the same state sequences (13-9-9) with “hypocrisy”. For example, “suck it up” and “racists and”. Another figure shows the buggy patterns, the buggy patterns identified by this RNN model include “...”, “!!!!”, etc.}
  \label{fig:pattern_interact}
\end{figure*}

\subsection{State Diagram View}
\label{subsec:state_diagram}

The \textit{State Diagram View} visualizes the finite state machine that is abstracted from the given RNN in the previous step. It provides an overview of the model behavior and helps users understand the semantic meanings of its hidden states.
Users can navigate through different state nodes to explore what prediction result a state often leads to and how many times this state has been visited. Specifically, each state is color-coded based on how frequently the input instances going through this state have  a specific prediction result. The size of a state node is determined by how many input sentences have visited this node during the training time. For example, in Fig.~\ref{fig:state_interact} \circled{A}, since the RNN model makes two possible predictions---positive comment or negative comment, all nodes are assigned to two distinct colors---\textcolor[HTML]{2196f3}{blue} for positive comments and \textcolor[HTML]{f50057}{red} for negative comments. Since there are fewer red nodes and the red nodes have a much smaller size than blue nodes, one can interpret that the training dataset has more positive comments than negative comments and the RNN model is more likely to make a positive prediction. 
The width of an edge between two states indicates how frequently this transition has occurred during the training time. The RNN model moves from one state to another state when it reads more words from a given input sentence. 

When a user hovers the mouse over a state, a tooltip is rendered to provide more details about this state, e.g., the number of training instances that go over this state and is eventually predicted to a specific result (Fig.~\ref{fig:state_interact} \circled{A}). When users click on a state, an information card (Fig.~\ref{fig:state_interact} \circled{B}) popped up showing the phrases and words that are frequently associated with this state in the training data. This feature allows users to interpret the semantic information memorized by hidden states. 
Clicking on a state also updates the instance view to filter out the input instances that do not visit this state during model prediction. 

\subsection{Pattern Summary View}
\label{subsec:pattern}

The \textit{Pattern Summary View} renders common patterns based on frequent state transitions during the training time. Basically, a frequent subsequence of states is viewed as a pattern, which can be further converted into a sequence of words based on the words or phrases associated with each state. {\tool} identifies two kinds of patterns: \textit{Influential Patterns} and \textit{Possible Buggy Patterns}. 

\textit{Influential Patterns} are patterns that change the model's intermediate predictions, e.g., changing from a positive comment to a negative comment after reading certain words in the middle of a given input sentence. To identify influential patterns learned from the training data, {\tool} first identifies the pivoting points (i.e., the point where the intermediate prediction changes) in the state trace of each training instance. Then {\tool} splits each state trace into subsequences based on the pivoting points. These subsequences are sorted based on their frequency and rendered in a descending order in the pattern summary view. 

\textit{Possible Buggy Patterns} are mined only from incorrectly predicted instances from the training data. These patterns indicate the cases where the RNN model does not learn well and thus makes a misprediction. To identify buggy patterns, {\tool} first divides the training data $S$ into two subsets according to the correctness of their prediction results. Let's denote the subset only include correct predictions as $S_c$ and the subset only include false predictions as $S_f$, respectively. Then we use TKS~\cite{fournier2013tks} (Top-K Sequential pattern mining) to mine frequent subsequence patterns from each subset. A subsequence pattern is considered possibly buggy if it only occurs in the misclassified inputs ($S_f$), not in the correctly classified inputs ($S_c$). These buggy patterns are sorted based on their frequency and rendered in a descending order in the pattern summary view.

Users can click on a specific pattern to see the top frequent phrases associated with this pattern (Fig.~\ref{fig:pattern_interact} \circled{A}). This pattern summary view allows users to know what patterns the model has learned, and how these patterns would affect the model's predictions. Furthermore, \textit{Possible Buggy Patterns} allows users to recognize potential prediction risks of an RNN model. Clicking on a pattern will update the instance view to filter out data instances that do not follow the selected pattern.

\subsection{Instance View}
\label{subsec:instance}

\begin{figure*}[t]
  \includegraphics[width=0.9\linewidth]{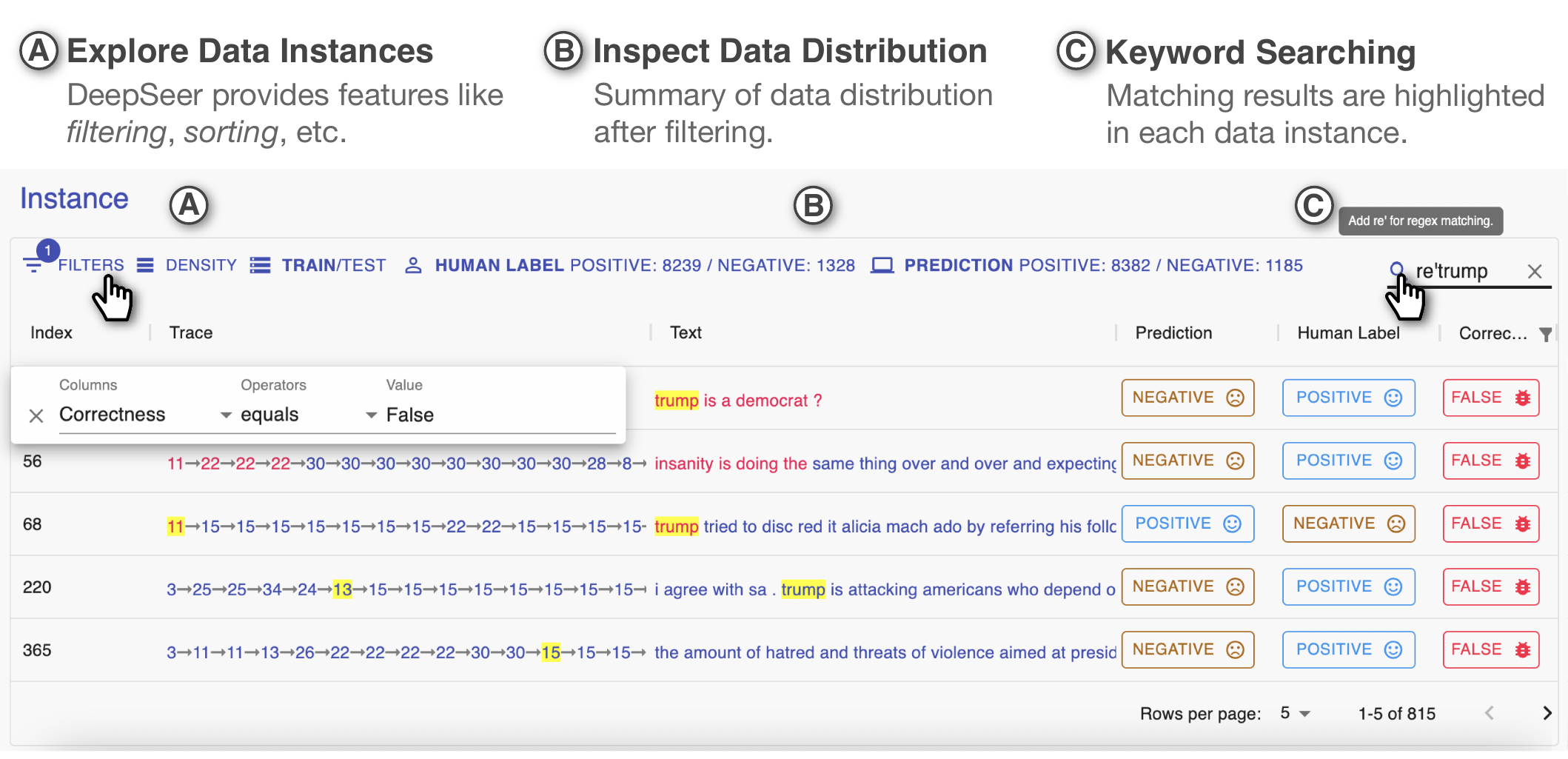}
  \Description{This figure shows the instance view and three different interactions with it. The instance view is a table showing the raw data, the column heads of this table are: index, trace, text, prediction, human label, and correctness. The first interaction with this instance view is that users can filter or sort the data in the instance view by clicking the “FILTERS” button on the top-left of this view. The second interaction is users can view the data distribution, which is right above the table. The third one is that users can search for specific words in this instance view by typing in the text box on the top-right corner of this view.}
  \caption{Interacting with the \textbf{\textit{Instance View}} in {\tool} to explore the data used to train/test an RNN model.}
  \label{fig:instance}
\end{figure*}

The \textit{Instance View} (Fig.~\ref{fig:instance}) is a scrollable data grid of the raw data used to train or test the model. The rows of the data grid are individual data instances, and the columns are: \textit{Index}, \textit{State Trace}, \textit{Text}, \textit{Prediction}, \textit{Human Label}, and \textit{Correctness} of the data instance. Users can sort and filter the data instances by each column (Fig.~\ref{fig:instance} \circled{A}). Users can use the \textcolor[HTML]{3f51b5}{TRAIN/TEST} tab to switch between training and test instances. The distributions of human labels and model prediction results are summarized and rendered on top of this view (Fig.~\ref{fig:instance} \circled{B}). As users filter the data instances, these distributions are also updated accordingly. Users can also search for specific input data  based on keywords or regular expressions (Fig.~\ref{fig:instance} \circled{C}). The matched results will be \textit{highlighted} for better visualization.  For each instance, its words and states are colored based on the intermediate prediction results. Clicking on a row in the instance view will update the state diagram view to render the state transitions of the selected data instance.

\subsection{Intermediate Prediction Results}
\label{subsec:intermediate}

\begin{figure*}[t]
  \includegraphics[width=0.85\linewidth]{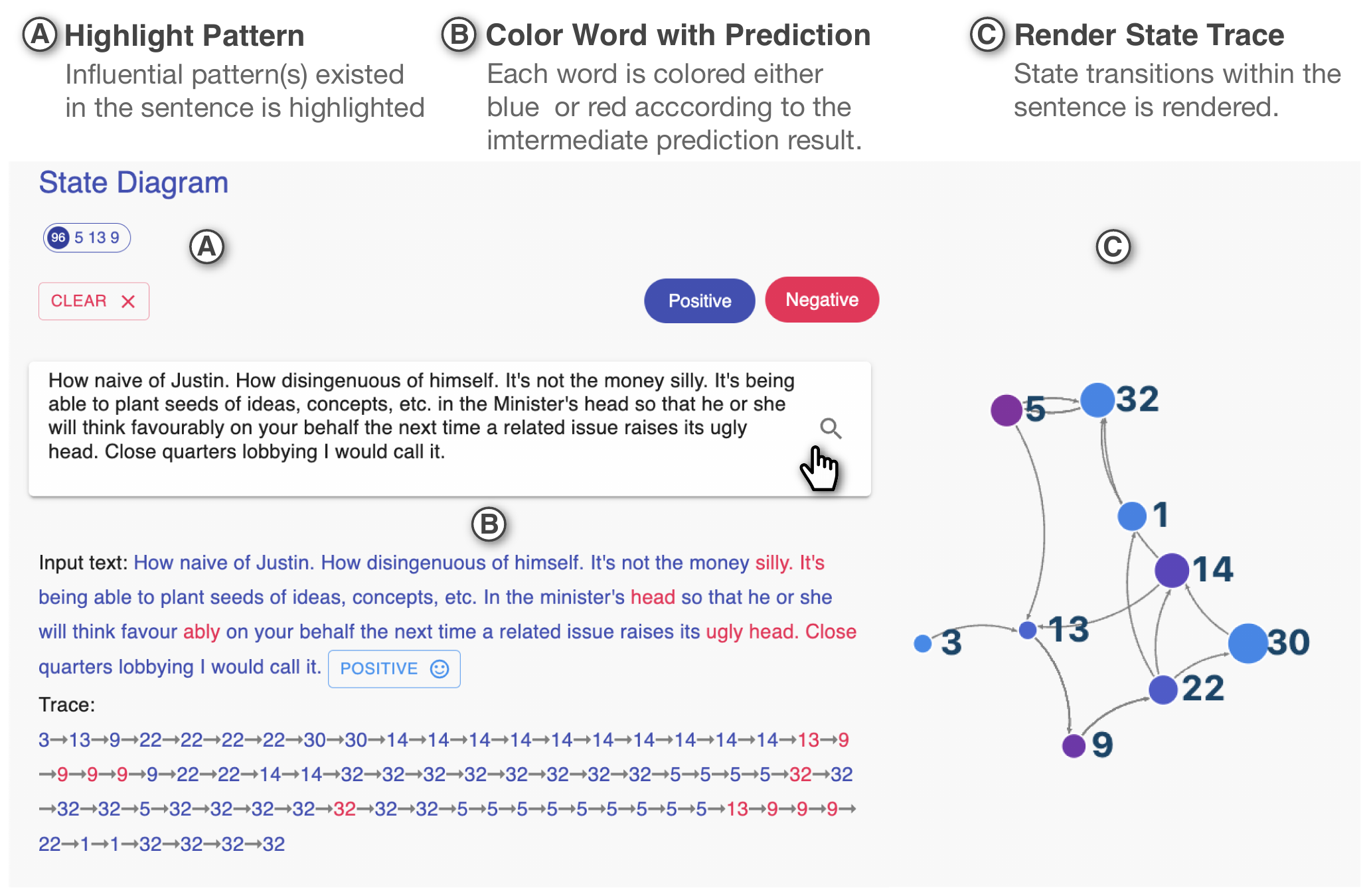}
  \Description{This figure shows an example of intermediate prediction results in DeepSeer. The top-left component of this figure is a text box, where users can type in a paragraph of text, then click on a magnifier to render the trace and intermediate prediction results of this text. The trace of the input text is plotted right to the text box. And the model’s prediction result is shown below the text box, along with the input text colored with different colors according to each word’s intermediate results}
  \caption{\textbf{\textit{Intermediate prediction results}} help users interpret model's prediction when debugging.}
  \label{fig:instance_explain}
\end{figure*}

The previous sections describe how users can use different views to achieve an overall understanding of the model behavior. Previous studies have shown that it is also important to allow users to inspect and debug model predictions on individual instances~\cite{ribeiro2016should,selvaraju2017grad,lundberg2017unified}. To support instance-level inspection and debugging, {\tool} allows users to enter an input sentence in the text box in Fig.~\ref{fig:instance_explain}. After clicking on the magnifier button, {\tool} renders the state trace of the model prediction on the given input. Furthermore, each word in the input sentence is colored based on the \textit{intermediate prediction result}. 
For example, in Fig.~\ref{fig:instance_explain} \circled{B}, ``\textcolor[HTML]{f44336}{red}'' and ``\textcolor[HTML]{3f51b5}{blue}'' indicate negative comments and positive comments respectively.
RNN typically uses the final hidden state after reading the entire sentence to compute the class probabilities. In \tool{}, the hidden state after reading each word in a sentence is fed into the output layer to generate intermediate predictions.
Through these intermediate predictions, users can inspect how the prediction result has changed over time as the RNN model reads more words in the input sentence. The pattern summary view is also updated with only influential patterns and possible buggy patterns related to the given input sentence. With these supports, users can quickly find suspicious words or phrases when debugging an incorrect model prediction.

%% file: scenario.tex
\section{Usage Scenario}
\label{sec:scenario}

Suppose Alice is a model developer, and she trains a toxicity detection RNN model using the Toxic dataset~\footnote{https://www.kaggle.com/c/jigsaw-unintended-bias-in-toxicity-classification}. This model predicts whether a sentence has a positive tone or negative tone. Her model achieves 95\% accuracy on the training data but only 89\% on the test data. Alice uses {\tool} to figure out why there is such a performance drift.  

\subsection{Visualizing and Understanding an RNN Model}

Alice first attends to the {\em state diagram view}, which gives her an overview of the trained RNN model as a finite state machine (Figure~\ref{fig:teaser} \circled{A}). In this state diagram, each node represents a group of similar hidden states in the RNN model. A \textcolor[HTML]{2196f3}{blue} node indicates that the model is more likely to give a \colorbox{primary}{\fontfamily{lmss}\selectfont \footnotesize \textcolor{white}{Positive}} intermediate prediction after visiting this state, while a \textcolor[HTML]{f50057}{red} node indicates that the model is more likely to give a \colorbox{secondary}{\fontfamily{lmss}\selectfont \footnotesize \textcolor{white}{Negative}} intermediate prediction. \responseline{Alice hovers her mouse over a red node named State 39}. As shown in Figure~\ref{fig:state_interact}, a tooltip then pops up showing that the model makes a negative intermediate prediction $39,948$ times while only $5,046$ times for positive ones, after visiting this state. 
When Alice clicks on the node of State 39, an information card is displayed on the right (Figure~\ref{fig:state_interact} \circled{A}), showing common words and phrases associated with the state, such as ``\textcolor[HTML]{3f51b5}{stupid}'', ``idiot'', and ``stupid and''. 
Alice glances over several sentences with these words in the {\em instance view} below (Figure~\ref{fig:teaser} \circled{C}) to check whether they are \emph{hate comments}. In this way, Alice confirms that her RNN model indeed learns some meaningful patterns from the training data. 

While inspecting text patterns associated with each state is helpful, Alice finds it cumbersome to check all states in the state diagram. So she switches to the {\em pattern summary view} (Fig.~\ref{fig:teaser} \circled{B}) to understand the model from another perspective. 
This view shows text patterns that have a significant impact on the model prediction (i.e., influential patterns). 
Alice finds some interesting patterns such as ``more fake news'' and ``stupid enough to''. When Alice clicks on one of the patterns, ``hypocrisy'' (Fig.~\ref{fig:pattern_interact} \circled{A}), it is expanded to show a list of other frequent patterns that are associated with the same state sequence, \circled{13} $\rightarrow$ \circled{9} $\rightarrow$ \circled{9}, sorted by frequency.\footnote{\responseref{} This single word, ``hypocrisy'', is associated with a sequence of three states, since it is tokenized into three tokens (hypo-, -cri-, -sy) in the training set, each of which is bound to one state. Such a tokenization mechanism is widely used in NLP to address out-of-vocabulary issues.} For example, this state sequence also memorizes ``suck it up'' (12 sentences), ``ignorant of'' (6 sentences), and ``suck and blow'' (4 sentences) in the training data. As Alice clicks on each pattern, the {\em instance view} is also updated to filter training and test data that does not follow the clicked pattern. Alice is also curious about which text patterns may cause incorrect model predictions. So she switches to the list of possible buggy patterns. These buggy patterns are summarized from misclassified sentences only, rather than the entire training dataset. Alice sees some patterns such as ``...'', ``!!!'', and ``???'' (Fig.~\ref{fig:pattern_interact} \circled{B}). It seems that her model learns some spurious correlations between punctuation and prediction results, which may have contributed to many errors. To prevent the model from learning these spurious correlations, Alice plans to remove this punctuation to clean the training data, which may lead to better model performance.

\subsection{Debugging an RNN Model}

Now Alice wants to dig into the data and investigates why some sentences are misclassified after having a high-level understanding of the model behavior. So she turns to the  {\em instance view} (Figure~\ref{fig:teaser} \circled{C}), which shows all training and test data in a paginated table. 

Alice first notices that the training data is not balanced. There are significantly more positive sentences (92062) than negative ones (7938). Alice then switches to the test data and finds misclassified sentences using the filtering feature on the {\em Correctness} column (Figure~\ref{fig:instance} \circled{A}). She copied a misclassified sentence to the text box and run the instance-level model explanation feature on it (Figure~\ref{fig:instance_explain}). Each word in the sentence is colored based on the intermediate prediction result. A state trace is also rendered below. 

Alice quickly notices a few words that her RNN model considers negative during the prediction, such as ``\textcolor[HTML]{f44336}{ugly head}.'' Even though such insulting words have been recognized by the model, this sentence is eventually predicted positive. It seems the model quickly forgets these insulting words after seeing the subsequent words in the sentence. For example, after seeing ``\textcolor[HTML]{3f51b5}{quarters}'', the intermediate prediction changes from negative to positive.

\begin{figure}[ht]
    \centering
    \vspace{-5pt}
    \includegraphics[width=0.9\linewidth]{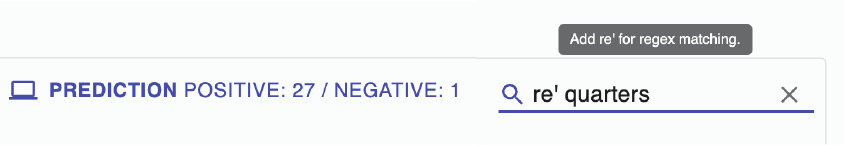}
    \Description{This figure shows the searching results of Alice. When she searches for “quarters”, there are 27 positive predictions while only 1 negative prediction on the training data.}
    \caption{Alice search for ``quarters'' in the training data.}
    \label{fig:quaters}
    \vspace{-5pt}
\end{figure}

To verify this hypothesis, Alice searches sentences that contain ``quarters'' in the training set using the keyword search feature in the {\em instance view}. Alice finds 27 positive sentences and only 1 negative one that contains ``quarters'' in the training set (Fig.~\ref{fig:quaters}). Since many sentences with ``quarters'' are positive, the model may have learnt a spurious correlation between ``quarters'' and the positive sentiment. Alice further confirms her hypothesis by looking at the corresponding state, \textcolor{primary}{State 22}, associated with the word ``quarters''. Alice finds that \textcolor{primary}{State 22} is a positive state. Therefore, Alice concludes that one reason for this misclassification is data imbalance, where most sentences with ``quarters'' are labelled as positive. \responseline{To address this, Alice believes that one possible solution is to collect more data with this keyword to balance her training set and then re-train her model.} Furthermore, given that the model quickly forgets an insulting word, Alice also plans to experiment with the long short-memory (LSTM) architecture with the attention mechanism, which can handle information in the memory for a longer time compared with a vanilla RNN. 

%% file: user-study.tex
\section{User Study}
\label{sec:user_study}

We conducted a between-subjects study with 28 participants to evaluate the effectiveness and usability of {\tool}. 
We used  LIME~\cite{ribeiro2016should}, a well-known tool for interpreting and debugging machine learning models, as a comparison baseline. \responseline{Though LIME is not specialized for RNNs, it is a widely-used tool to understand and debug models. It has 10.3K stars and 1.7K forks on GitHub~\footnote{https://github.com/marcotcr/lime}, and its Python package has been downloaded 16M times on PyPI~\footnote{https://pepy.tech/project/lime}. Therefore we choose it as a more realistic baseline.} Given a model prediction, LIME can generate an explanation with importance scores for elements in the input data (e.g., words in an input sentence).
To enable a fair comparison, we built an interface for LIME similar to \tool{}. The interface includes the existing visualizations provided by LIME and also includes the Instance View as in \tool{}. It does not include the state diagram view and the pattern summary view, which are the novel contributions of \tool{}. \responseline{We investigated the following research questions to assess the overall usefulness of {\tool} compared with LIME:}

{
\responseref{}
\begin{itemize}[leftmargin=*]
    
    \item RQ1: To what extent does {\tool} enhance users' understanding of an RNN model compared with a commonly used model explanation and debugging tool?
    
    \item RQ2: To what extent does {\tool} improve the accuracy of identifying the root cause of a misprediction of an RNN model compared with a commonly used model explanation and debugging tool?
    
\end{itemize}
}

\subsection{Participants}

We recruited 28 participants (5 female and 23 male) through several graduate student mailing lists of the CS department and the ECE department at the University of Alberta.\footnote{This human-participated study is approved by the university's research ethics office.} 
All participants had at least basic machine learning background. 15 participants were Ph.D. students, and the rest were Master's students. 
23 participants had 2-5 years of machine learning experience, 3 participants had more than 5 years, and 2 participants had about 1 year. Regarding their RNN experience, 9 participants had more than 2 years of experience, 7 participants had 1 year, and 12 participants had less than 1 year. Participants also self-reported their familiarity with developing RNN models in a 7-point Likert scale question. The median is 5, with 1 referring to ``{\em I have only heard about RNNs but never used it}'' and 7 referring to ``{\em I'm able to build an RNN model by myself}.'' 25 participants said they had not used any debugging tools for DL, while 3 participants said they have used Tensorboard~\cite{tensorflow2015-whitepaper}. The studies were conducted on Zoom. Both {\tool} and LIME were deployed as web applications that participants could access from their personal computers. 

\subsection{RNN Models}
\label{subsec:tasks}

Since {\tool} is designed for visualizing and debugging RNN models, we trained two RNN models for two popular ML tasks. \responseline{For each RNN model, the dimension of a hidden state vector is 256.} The first ML task is to predict whether a question asked on Quora is sincere or insincere. It is originally from a featured competition from Kaggle, a popular online machine learning and data science community~\cite{quora}. In this task, our RNN model is trained on 100,000 Quora questions, each of which is labeled as sincere or insincere. \responseline{The training accuracy of this RNN model is 93.93\%, and the test accuracy is 89.07\%.} The second ML task is to predict the topic of a news article from a news corpus called AG's News~\cite{agnews}. This task is a well-known benchmark for topic classification research~\cite{zhang2015character}. In this task, our RNN model is trained on 109,886 news articles labeled into four news topics, including ``Sports'', ``Business'', ``World'', and ``Science and Technology.'' \responseline{The training accuracy of this RNN model is 91.57\%, and the test accuracy is 87.68\%.} \tool{} abstracts each RNN model into 40 states. This number is decided empirically to achieve a good balance between accuracy and the cognitive effort of inspecting a state diagram. \responseline{We further provide a faithfulness analysis of the abstracted model in Appendix~\ref{appendix:states}}. During a user study session, we randomly assigned one of the two RNN models to a participant to finish the model understanding and debugging tasks. Interface of {\tool} for each task can be found in Appendix~\ref{appendix:interface}.

\begin{figure*}[t]
    \centering
    \includegraphics[width=0.95\linewidth]{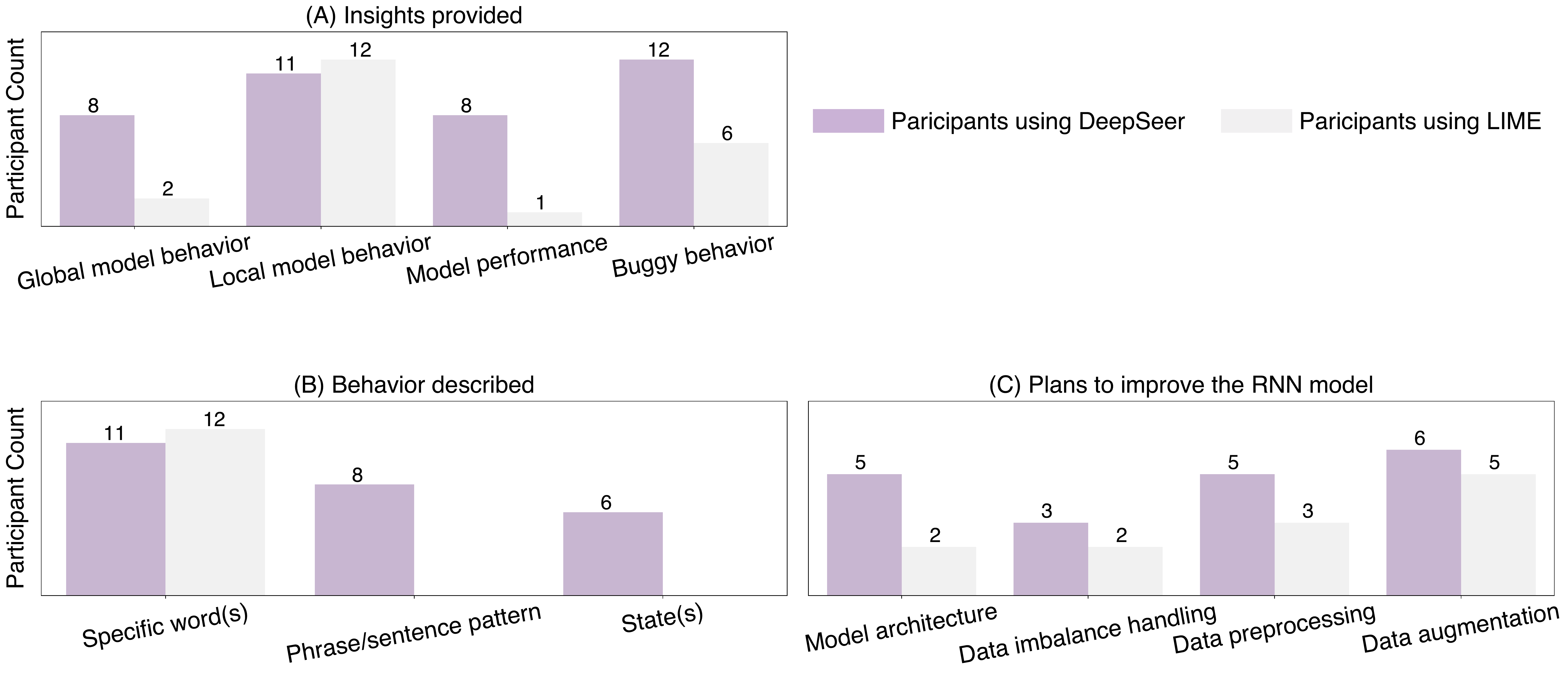}
    \Description{This figure shows the participants’ performance on model understanding. It shows that compared with LIME users, DeepSeer users (a) provided more diverse insights, (b) were more likely to use deductive reasoning methods, and (c) came up with more actionable plans for improving the RNN model.}
    \caption{\responseref{} Participants' performance on model understanding. (A) The number of participants who mention different aspects of model behavior. (B) The number of participants who describe model behavior in different ways. (C) The number of participants who make different suggestions on model improvement.}
    \label{fig:t1}
\end{figure*}

\subsection{Protocol}

\responseline{We design a between-subjects user study where users experience one condition and one RNN model in each study session. We choose a between-subjects design rather than a within-subjects design since experiencing one condition takes around 60 minutes. Experiencing two conditions in a within-subjects design would require 120 minutes, which is too long and can lead to significant fatigue and frustration.}
At the beginning of each session, we asked the participants for their permission for recording. Given that this was a between-subjects study, each participant was assigned to only one RNN model in one condition. In each session, participants were only allowed to use the given tool: {\tool} in the experiment condition or LIME~\cite{ribeiro2016should} in the control condition. The assigned RNN models and conditions were counterbalanced across participants.
At the beginning of each study session, participants were asked to first watch a 5-min tutorial video of the 
assigned tool, and then spend 5 minutes familiarizing themselves with the tool. Then, participants were  given 30 minutes to use the assigned tool to explore an assigned RNN model and share their understanding of the model behavior through a questionnaire. The questionnaire included three main questions: (1) {\em What insights have you got about the model's performance and behavior?} (2) {\em Did you find any bugs or limitations of the RNN model? If yes, what kind of bugs have you found?} (3) {\em How will you further improve the model?} After filling out the model understanding questionnaire, participants were given 10 minutes to debug 5 incorrect model predictions using the assigned tool. For each incorrect prediction, they were asked to write down why the input data was misclassified and submit their answers through a questionnaire. At the end of the study session, participants were asked to fill out a survey to share their experiences. In particular, the post-study survey included the NASA Task Load Index (TLX) questions~\cite{hart1988development} to measure the cognitive load of the study. Each participant received a \$25 Amazon gift card as compensation for their time.

%% file: results.tex
\section{User Study Results}
\label{sec:results}

This section describes the results of the between-subjects user study. \responseline{We first present and analyze participants' performance differences on model understanding and debugging tasks when using {\tool} and LIME. Then we present participants' perception on {\tool}'s tool features as well as cognitive load.} For brevity, we use P1-P14 to denote the participants using {\tool}, and P15-P28 to denote the participants using LIME~\cite{ribeiro2016should}.

\subsection{RQ1: User Performance on Model Understanding}
\label{sec:understanding}

{\responseref{}

To evaluate user performance on model understanding, two authors manually assessed and coded participants' responses and counted the number of correct insights about model behavior shared by participants. Specifically, these two authors had 4 meetings to develop a codebook and resolve labeling inconsistencies. Eventually, 651 codes were generated and categorized into 32 themes. The final Cohen's Kappa score is 0.9061. Note that one insight is considered correct only if both two authors agree.

Overall, participants using {\tool} provided more insights (53 vs.~21) than participants using LIME. The mean difference of insights provided per participant (2.3) is statistically significant (Welch's t-test: $p=0.0003$). Fig.~\ref{fig:t1}~provides a breakdown of different kinds of insights shared by participants. Participants using {\tool} shared much more insights about {\em global model behavior}, {\em model performance}, and {\em buggy behavior}. For instance, P12 said, ``\textit{It looks like the model is placing a lot of weight in the latter half of an input sentence.}'' P9 wrote, ``\textit{this model is often confused by the Business and Science categories.}'' Furthermore, participants using {\tool} often referred to text patterns and states when describing model behavior, while participants using LIME mostly referred to specific keywords. P25 said, ``\textit{it is not easy to summarize patterns [with LIME] when the size of dataset is large and there are many classes.}'' \camerareadyrevision{Since LIME is designed for local explanations, it is not surprising that only 2 participants using LIME were able to derive global explanations for an assigned RNN model.}

%----------------------------------------
\begin{table*}[t]
\small
\centering
\caption{The average of responses shared by participants in the model debugging task.}
\begin{tabular}{l| c c | c c | c c}
\toprule
\multirow{2}{10em}{} & \multicolumn{2}{c|}{Quora} & \multicolumn{2}{c|}{AGNews} & \multicolumn{2}{c}{Overall}  \\
\cline{2-7}
 & {\tool} & LIME & {\tool} & LIME & {\tool} & LIME \\
\hline
Reasonable explanations provided per participant & 4.3 & 2.7 & 4.3 & 1.1 & 4.3 & 1.9 \\
Fault-inducing keywords mentioned per participant & 5.3 & 2.7 & 4.6 & 1.4 & 4.9 & 2.1 \\
Non-fault-inducing keywords mentioned per participant & 0.8 & 3.7 & 1.0 & 3.3 & 0.9 & 3.5\\
\bottomrule
\end{tabular}
\label{tab:t2}
\end{table*}
%----------------------------------------

Participants using {\tool} also provided more useful and diverse suggestions about model improvement compared with LIME users (Fig.~\ref{fig:t1}~(C)). 
Note that a suggestion is considered useful if it is related to the root causes of observed model misprediction and is accepted as an effective model improvement mechanism in the ML community. In particular, 5 participants using {\tool} noticed the error pattern of forgetting previous tokens after reading more tokens and suggested adding an attention layer, while only 2 participants using LIME noticed this. P10 wrote, ``\textit{For many false predictions, the model is likely to give the right prediction at the beginning, but then turns to the wrong direction. Probably we could reduce the length of temporal dependencies with something like the attention mechanism.}'' 
\responseline{Finally, participants using {\tool} spent 27~min 53~s ($\sigma = $ 2~min 44~s) on average, while participants using LIME spent 28~min 19~s ($\sigma = $ 3~min 15~s). We do not observe a significant difference in task completion time.} 

}

\subsection{RQ2: User Performance on Model Debugging}
\label{sec:debugging}

To evaluate user performance on the model debugging task, we counted the number of reasonable explanations provided by participants over five misclassified sentences. To assess the correctness of participants' answers, two authors first manually inspected the hidden states of the RNN and also the training data to diagnose the five misclassifications. Their investigation results were used as the ground-truth misclassification explanations. Then, they checked whether the participants' explanations were consistent with the ground truth. 

As shown in Table~\ref{tab:t2}, participants using {\tool} provided more reasonable explanations for misclassification. 
Participants using {\tool} provided 4.3 reasonable explanations for 5 misclassified sentences on average, while participants using LIME only provided 1.9 reasonable explanations. The mean difference of 2.4 is statistically significant (Welch's t-test, $p<0.0001$).

Furthermore, we counted the number of correct fault-inducing keywords mentioned by participants. 
Participants using {\tool} identified more correct fault-inducing keywords than those using LIME (mean: 4.9 vs. 2.1). The mean difference of 2.8 is statistically significant (Welch's t-test, $p<0.0001$). In addition, participants using LIME misrecognized more keywords (mean: 3.5 vs. 0.9) as fault-inducing keywords (Welch's t-test, $p<0.0001$). This is because LIME first learns a surrogate sparse linear model to simulate the RNN model and then computes word importance based on the linear model. This sometimes leads to unreliable explanations. Some participants also noticed this during the study. 
P22 commented, ``\textit{in some cases, I found that the tool [LIME] did not generate a reliable explanation.}''

One interesting observation is that participants using {\tool} were capable of identifying more complex error patterns beyond word patterns. For example, P9 answered, ``\textit{At the very beginning, `football' indicates the model to predict sports, which is exactly what the model does. But when `UK' appears, the state transits to `world'  [related state] and got stuck there.}'' None of the LIME users provided such insights, since LIME treats individual words separately and cannot capture the dynamics of the model's decision process. 

Finally, participants using {\tool} completed this task in an average of 9~min 9~s ($\sigma = $ 1~min 21~s), while participants using LIME took an average of 9~min 25~s ($\sigma = $ 0~min 11~s). There is no significant difference in task completion time.

% \iffalse
\subsection{User Perception and Cognitive Load}
\label{subsec:user perception}

Our post-study survey solicited participants’ feedback on all key features of {\tool}. Overall, participants considered {\tool}’s visual encoding and interface intuitive, helpful, and clear. 
Among 14 participants, 13 of them self-reported that they would like to use {\tool} when developing and debugging RNN models in the future, while 1 participant stayed neutral. The median is 6.5 on a 7-point Likert scale (1---I don't want to use it at all, 7---I will definitely use it if available). We report participants' qualitative feedback on the key features of {\tool} from both post-study survey and user study recordings below.

\begin{figure*}[t]
    \centering
    \includegraphics[width=\linewidth]{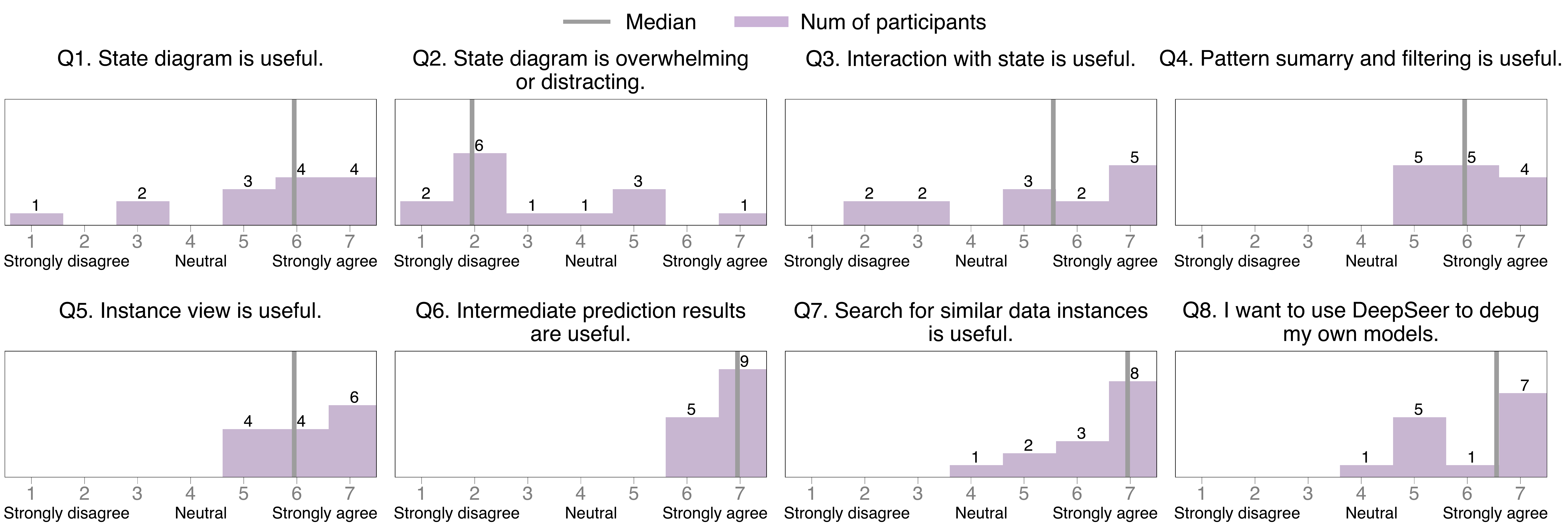}
    \caption{Participants' ratings about {\tool}'s tool features (1 is ``strongly disagree'' and 7 is ``strongly agree'')}
    \Description{This figure shows the participants’ rating of tool features of DeepSeer on a 7-point Likert scale (1 means strongly disagree and 7 means strongly agree). It shows that all of our tool features are highly appreciated by the participants, especially the state diagram, the instance view, and the intermediate prediction results. The median rating of “the state diagram is useful” is 6. The median rating about “the instance view is useful“ is also 6, while the median rating for “the search feature in the instance view is useful” is 7. The median rating of “the intermediate prediction results are useful” is 6. And 13 out of 14 participants also agreed that they want to use DeepSeer to debug their own models.}
    \label{fig:feature rating}
\end{figure*}

\vspace{1mm}

{\bf\em \noindent Intermediate Prediction Results.} 
\sloppy All 14 participants using {\tool} found the on-demand intermediate prediction results provided by {\tool} useful. The median rating is 7 out of 7. 
\responseline{P9 mentioned, ``\textit{stepping through intermediate predictions help me understand why model makes a wrong prediction. For example, the text is apparently about sports. However, the model goes into state 20 which is not quite related to sports.}''}
Moreover, participants also liked the color-coding of intermediate prediction results. 

\vspace{1mm}

{\bf\em \noindent State Diagram.} Among 14 participants,  11 of them considered the state diagram in {\tool} useful. The median rating is 6. 
\responseline{P12 commented, ``\textit{extracting an RNN model as a state diagram is nice, and I think it will also be helpful when interpreting [RNN models] with more complex data such as medical data.}''}
While the majority of participants did not find the state diagram overwhelming, 3 found it slightly overwhelming and 1 found it very overwhelming. 7 out of 14 participants found it useful to interact with the state diagram, e.g., seeing statistical distribution over states and keywords associated with each state. 
\vspace{1mm}

{\bf\em \noindent Pattern Summaries.} 9 out of 14 participants found that seeing the patterns in the pattern summary view and filtering the dataset based on a specific pattern are useful (median rating: 6). P2 mentioned, ``\textit{it is good for us to see inside of the model and find the bug with possibly buggy patterns.}'' 
In particular, participants also mentioned that the pattern summary view is helpful for debugging. 
\responseline{P8 said, ``\textit{I can click the buggy patterns to check related sentences. This helps me identify why model usually mis-classify [sentences] with these patterns.}''}

\vspace{1mm}

{\bf\em \noindent Searching and Filtering Instances.} Most participants agreed that it is useful to interact with each data instance in the instance view (median rating: 7) and search for similar instances (median rating: 7). P4 mentioned, ``\textit{I like the colors associated with each label, I feel this helped a lot with looking at examples. I also liked how it clearly showed examples with their true class and prediction. Also, the ability to also filter examples by correctness, prediction, and the true label was very helpful for me.}''

\vspace{1mm}

{\bf\em \noindent Limitations and Suggestions.} 5 out of 14 participants pointed out that it would be better if {\tool} could provide additional statistical information about model accuracy, \eg, confusion matrix. 
1 participant suggested that adding the confidence score for the explanation could help them make more informed decisions. 
1 participant found the state diagram mentally demanding. P6 said,  ``\textit{state diagram seemed a bit hard to interpret by just looking at it and probably wouldn't be immediately intuitive to a user opening this application up initially.}''

\vspace{1mm}

\begin{figure}[t]
    \centering
    \includegraphics[width=\linewidth]{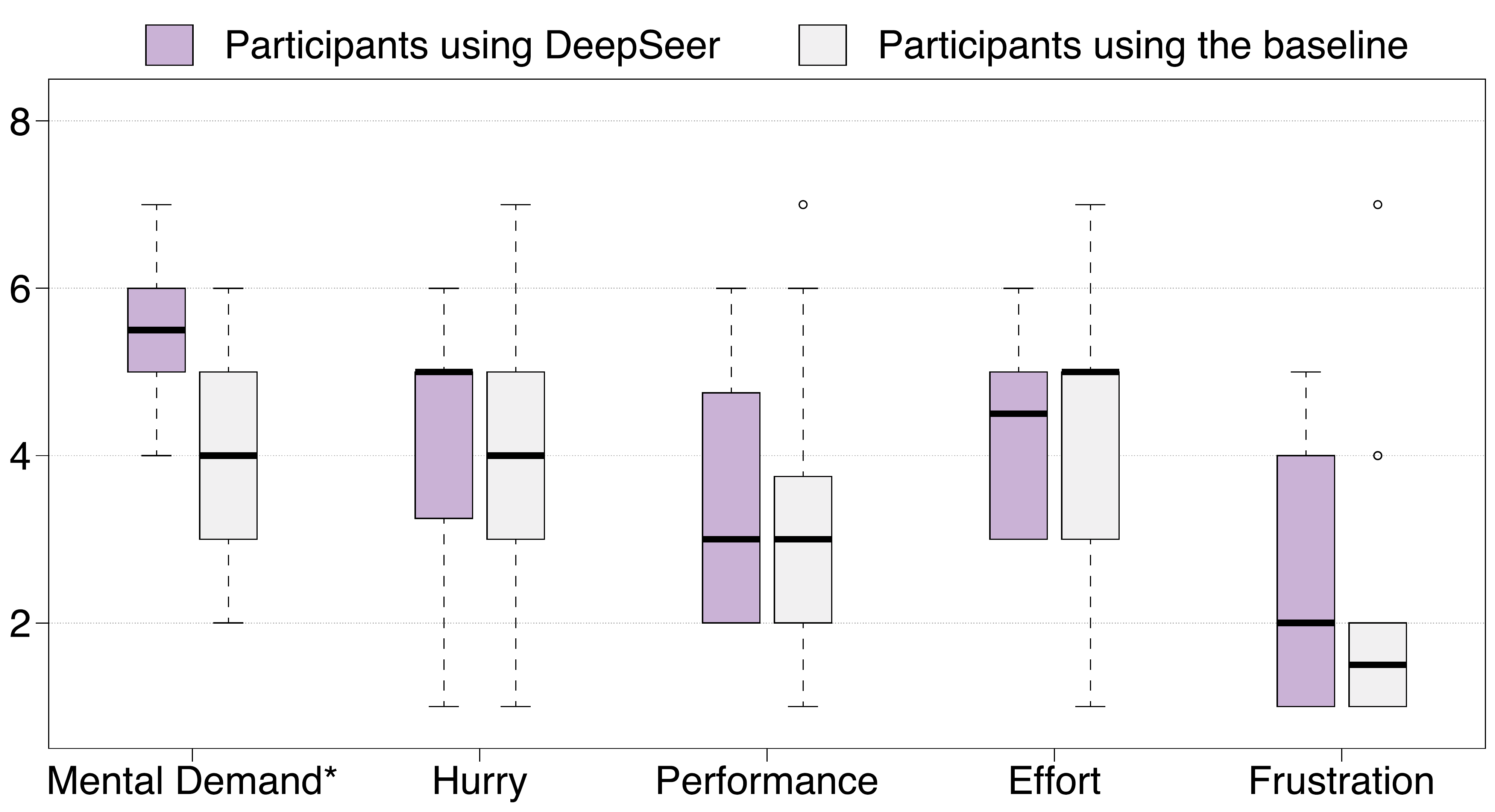}
    \Description{This figure shows the cognitive load of participants when using DeepSeer versus LIME. The cognitive load is measured in five different dimensions, including mental demand, how hurry participants feel during the study, participants’ self-assessment of their performance, participants’ self-assessment of the effort needed to complete the task, and how frustrated participants feel. This figure shows that participants using DeepSeer have a higher cognitive load in terms of mental demand (median rating 5 versus 4). However, DeepSeer users’ cognitive loads in other dimensions are similar to LIME users’.}
    \caption{Participants' ratings about cognitive load (1 means ``strongly disagree'' and 7 means``strongly agree'', * means the mean difference is statistically significant.)}
    \label{fig:experience}
\end{figure}

\noindent{\bf\em Cognitive Overhead.} In the post-study survey, participants rated the cognitive load of the study via the NASA TLX questionnaire~\cite{hart1988development}.  Fig.~\ref{fig:experience} shows their ratings for the five NASA TLX questions. We found that there was no significant difference when using {\tool} vs.~LIME in terms of hurry, performance, effort, and frustration (Welch's t-test: $p=0.7731$, $p=0.7244$, $p=0.6916$, and $p=0.5620$). Since {\tool} renders much more information about model behavior (e.g., a state diagram, a pattern view, on-demand intermediate prediction results), participants using {\tool} felt more mental demand (median value: 5 vs.~4, Welch's t-test: $p=0.0011$). 

%% file: discussion.tex
\section{Discussion}
\label{sec:discussions}

\subsection{Design Implications}

The user study results suggest that {\tool} helps users achieve a more comprehensive understanding of the assigned model, and perform better on model debugging compared with the baseline tool, LIME~\cite{ribeiro2016should}. We believe this is largely attributed to {\tool}'s interactive support for explaining the model's global and local behavior. While a few studies have discussed about the importance of global and local explanations~\cite{montavon2018methods, hohman2019gamut}, our work provides specific insights on how to support global and local explanations in a unified interface for RNN models. 

In {\tool}, global explanations are mainly rendered in the \textit{State Diagram View} and the \textit{Pattern Summary View}. The abstracted state diagram helps users interpret the hidden states and complex transitions among these states, while the summarized text patterns help users quickly identify either influential or buggy patterns learned by the model. These global explanations boost users' understanding and debugging process. Despite all the benefits of global explanations, we found it still necessary for participants to have the instance-level explanation to contextualize their understanding of model behavior. In particular, given a specific state or text pattern, user study participants often got curious about how it sounds in different texts. In the post-study survey, they highly appreciated the \textit{Intermediate Prediction Results} feature. {\tool} allows users to zoom into local explanations by actively filtering instances based on selected states or patterns, as well as zooming back to the model's global behavior by tracing back to the state diagram. 
Through these ways, global and local explanations are served as a synergistic loop for model understanding and debugging.

Furthermore, we find that users cared about how the given explanations are derived from the internal decision-making process of an RNN model. When using LIME~\cite{ribeiro2016should}, 4 out of 14 participants using LIME questioned the explanations (highlighted keywords) given by LIME. For instance, P24 commented in the post-study survey, \textit{"I hope LIME can provide a reason why some words have a high sincere or insincere score."}
As a more tangible and actionable solution, {\tool} not only communicates the correlation between specific features in an input to a prediction result, but also communicates the internal decision-making process of a model. {\tool} renders model's decision-making process in two ways. First, it renders the transition between different internal states of the model in the state diagram view. Second, for an individual prediction, it renders the intermediate prediction results as well as their correspondence to the internal states of a model. By inspecting such a decision-making process, users can better understand how the model arrives at a specific prediction and gain more trust from the generated explanations.

As an interactive XAI tool, it is also important to provide users with interpretable explanations, especially for RNN models. Note that a few prior techniques have tried to visualize the decision process of RNN~\cite{karpathy2015visualizing,strobelt2017lstmvis}. However, they usually only directly visualize the value of each hidden state. For instance, LSTMVis~\cite{strobelt2017lstmvis} visualizes the change of hidden state values in parallel coordinates. Given that hidden state values are essentially numerical values in a high-dimensional space, it is challenging to interpret their semantic meanings. To address this challenge, {\tool} bundles hidden states with associated words and phrases in a text corpus and visualizes the transition between them as a state diagram. In this way, the internal decision-making process becomes more interpretable to non-experts.

{\responseref{}

\subsection{Target Users and User Expertise}

{\tool} is designed for any developers who needs to train and debug an RNN model by themselves. They can be experienced ML developers, regular software developers who just started learning RNNs, or students who use RNN in a course project. In the user study, we recruited participants with diverse expertise in RNN, including 4 participants with less than 1 year of RNN experience, 5 with 1 year, 3 with 2---5 years, and 2 with more than 5 years. Our further analysis shows that, while participants with more RNN experience performed slightly better, the difference was not significant (Fig.~\ref{fig:rnn_expertise}). This implies the effectiveness of f{\tool} is not strongly correlated to their expertise. 

\begin{figure}[t]
    \centering
    \includegraphics[width=\linewidth]{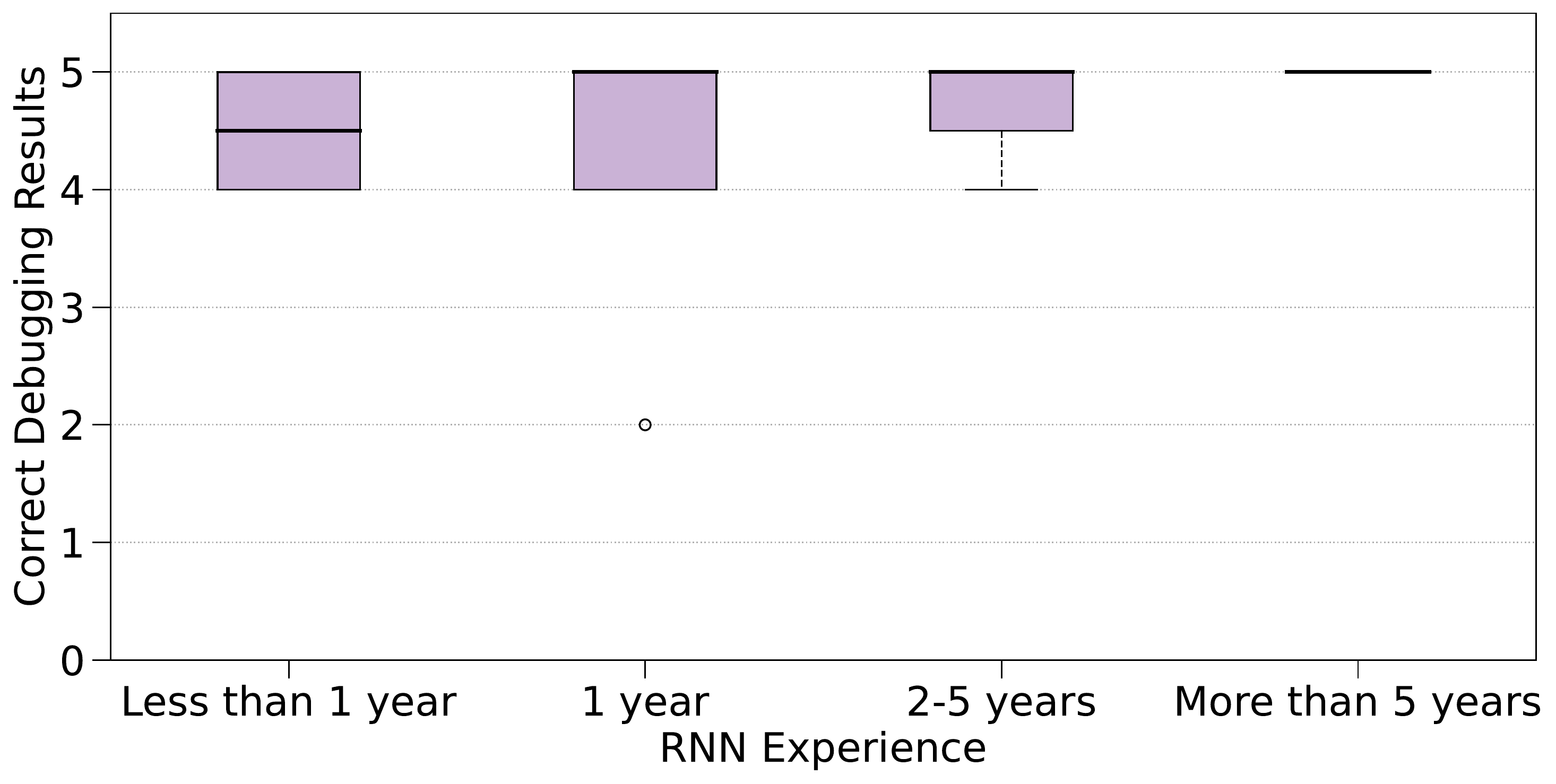}
    \caption{{\tool} users' performance on model debugging with various levels of RNN expertise.}
    \Description{This figure shows DeepSeer users’ performance on model debugging with various levels of RNN expertise. The median performance of four different RNN experience groups (less than one year, one year, two to five years, and more than five years) are 4.5, 4, 5, and 5, respectively. There is no significant performance difference among different groups.}
    \label{fig:rnn_expertise}
\end{figure}

}

\subsection{Generalization to Different ML Tasks and Models}

Though our work has only evaluated {\tool} on sentiment analysis and topic modeling tasks, we believe {\tool} can generalize to different NLP Tasks as well. To reuse {\tool} for other tasks, one may  consider adapting the color mapping mechanism for abstract states. For example, for machine translation tasks, one can color each state according to the part-of-speech tag. By inspecting each state's color and associated words, users could interpret if an RNN model is translating a sentence correctly.

In this work, we focused on RNNs, which is a representative model architecture for processing sequential data. In addition to RNNs, it may be possible to use {\tool} to interpret RNN variants such as Bidirectional-LSTM~\cite{liu2019bidirectional} or Transformers~\cite{vaswani2017attention}. While the principle of Bidirectional-LSTM is similar to a naive RNN, some adaptions to the state abstraction method are required. For instance, one should consider collecting the model's hidden states when processing the input text in both two directions.
Since transformers are permutation-invariant, they process all words in an input sentence at the same time, not sequentially. Therefore, we can no longer bundle a word with a hidden state. However, one can treat the output of each hidden layer as a concrete state. Then the transition can be built among different hidden layers instead of among different words.

\subsection{Limitations and Future Work}

\camerareadyrevision{One limitation of our user study design is that the comparison baseline, LIME~\cite{ribeiro2016should}, is designed for generating local explanations instead of global explanations. Thus, we cannot directly compare the global explanation effectiveness of {\tool} to LIME.} Besides, LIME is not specialized for RNN. While there are RNN-specific tools, such as LSTMVis~\cite{strobelt2017lstmvis} and RNNVis~\cite{ming2017understanding}, we failed to run them on our RNN models after several attempts due to version compatibility issues. Both LSTMVis and RNNVis were built years ago on out-of-dated DL frameworks (TensorFlow r0.12 and Torch~7), which can no longer be used to analyze DL models built by later framework versions. Significant efforts are needed to re-implement them. Therefore, we consider such re-implementation out of the scope of this work. An alternative solution can be creating variants of {\tool} by disabling some key features as baselines, which can help us better attribute the success of {\tool} to individual features. This is worth investigating in future work.

\responseline{As we are researchers from an R1 university, we do not have access to professional developers and data scientists who build RNN models in their work. Instead, we recruited graduate students who have experience in building RNN models. ML practitioners may share more interesting insights compared with graduate students. }

Additionally, our user study has only evaluated {\tool} on RNNs for sentiment analysis and topic classification. To comprehensively investigate the usefulness of {\tool}, one can consider leveraging {\tool} to understand and debug RNN models for other tasks, beyond text classification, \eg, machine translation. 

\responseline{Furthermore, our current design only supports visualizing and analyzing one RNN model. Once the bugs are identified with the help of {\tool}, re-training the RNN model is needed. Therefore, a possible future direction is to develop tool support for comparing two or more versions of an RNN model~\cite{murugesan2019deepcompare}. Besides, one can also improve {\tool} by designing tool support for model tracking~\cite{amershi2015modeltracker} and model selection~\cite{bogl2013visual}.}

%% file: conclusion.tex
\section{Conclusion}
\label{sec:conclusion}

In this paper, we present a novel system called {\tool} to help ML developers understand and debug recurrent neural networks. {\tool} makes use of a state abstraction method that bundles semantically similar hidden states of an RNN model and abstracts it to a finite state machine. Through {\tool}, users can explore both the model's global and local behavior, and also debug incorrect predictions. We demonstrate {\tool}'s usefulness and usability through a between-subjects user study with 28 model developers on two different RNN models. The results show that {\tool}’s tightly-coordinated views brought developers a deeper understanding of an RNN model compared with a popular XAI technique, LIME. Furthermore, participants using {\tool} were able to identify the root causes of more incorrect predictions and provide more actionable plans to improve the RNN model. In the end, we discuss the design implications from {\tool}, and propose several promising future work directions.

%% file: appendix.tex
\newpage 

\appendix

\setcounter{figure}{0}
\def\thefigure{\Alph{section}\arabic{figure}}

\setcounter{table}{0}
\def\thetable{\Alph{section}\arabic{table}}

\def\thealgocf{\Alph{section}\arabic{algocf}}

{\responseref{}
\section{State Abstraction}
\label{appendix:state_abstraction}

In this section, we present the technical details of state abstraction for an RNN model. The core idea is to extract all possible hidden states of an RNN model using training data and then group similar hidden states to build a finite state machine (FSM). Algorithm~\ref{alg:abstraction} shows the procedure of state abstraction of an RNN model. 

\begin{algorithm}[h]\responseref{}
\caption{\responseref{}State abstraction of an RNN model.}
\label{alg:abstraction}
\KwIn{a trained RNN model $R$, training data $\mathcal{X}_{train}$, PCA dimension $k$, number of states $n$}
\KwOut{PCA model $P$, GMM model $G$, an abstraction model $A=\{P, G\}$}
$\mathcal{H}~\gets~\{\}$\;
\For{$x~$in$~\mathcal{X}_{train}$}{
$H~\gets~\mathsf{record\_hidden\_states}(R, x)$\;
$\mathcal{H}$.append$(H)$\;
}
$P~=~\mathsf{PCA}(\mathcal{H}, k)$\;
$G~=~\mathsf{GMM}(P(\mathcal{H}), n)$\;
\KwRet{$A$}\;
\end{algorithm}

Algorithm~\ref{alg:abstraction} takes two inputs: a trained RNN model $R$ and training data $X_{train}$, as well as two parameters: PCA (principal component analysis) dimension $p$ and number of abstracted states $n$. Given a trained RNN model, we first iterate through all the training data $\mathcal{X}_{train}$ to collect all possible hidden states $\mathcal{H}$ from the RNN model (Line 1:5). Line 3 records all the hidden states $H$ in an RNN model when processing a specific input instance $x$. Suppose an input instance $x$ has $l$ words, then $l$ different hidden states will be produced when the RNN model processes these $l$ words sequentially. For example, given the sentence ``I love Machine Learning'', the RNN model will process four words: ``I'', ``love'', ``machine'', and ``learning'' sequentially. Therefore, four different hidden state vectors will be produced and recorded when the RNN model processes this sentence.

After recording all the hidden states using the training data, we create a PCA model $k$ to reduce the dimension of these hidden states into $p$ (Line 6). Meanwhile, we also obtain a PCA model $P$. Now we abstract $|P(\mathcal{H})|$ dimension reduced hidden states into $n$ abstracted states. Different from the prior work~\cite{du2019deepstellar}, which uses a grid-based method, we adopt a GMM (Gaussian mixture model~\cite{mclachlan1988mixture}) $G$ to cluster these dimension-reduced hidden states (Line 7). After executing Line 7, we obtain a GMM model $G$. Our abstracted model $As$ has now been built, which consists of a PCA model $P$ and a GMM model $G$. Note that both PCA and GMM are implemented with scikit-learn with default parameters except ``$\mathsf{n\_components}$''.

In our usage scenario and user study, we set the PCA dimension $k$ as 20 and the number of states $n$ as 40. We further show that the abstraction model using this setting can provide consistent predictions compared with the original RNN model in Appendix~\ref{appendix:faithfulness}.

}

{\responseref{}

\section{Faithfulness of State Abstraction}
\label{appendix:faithfulness}

\begin{table*}[t]
    \responseref{}
    \centering
    \caption{\responseref{} Quantitative comparison between the abstraction model predictions to the original RNN's predictions.}
    \begin{tabular}{c|cc}
         \toprule
         & Prediction consistency on training set & Prediction consistency on test set \\
         \hline
         Toxic & 99.88\% & 99.88\% \\
         Quora & 97.30\% & 97.04\% \\
         AGNews & 86.28\% & 85.59\% \\
         \bottomrule
    \end{tabular}
    \label{tab:faithfulness}
\end{table*}

In this section, we show that the abstracted model (i.e., the finite state machine) can make consistent predictions as the original RNN model in three different tasks, one from the usage scenario (Section~\ref{sec:scenario}) and the other two from the user study (Section~\ref{subsec:tasks}). We measure the prediction consistency between the finite state machine and the original RNN. Suppose the dataset is $X=\{x_1,x_2,\dots,x_N\}$, the abstracted model is $\mathcal{F}$, and the RNN model is $\mathcal{R}$. The prediction consistency can be calculated through Eq.~\ref{eq:consistency}.

\begin{equation}
\label{eq:consistency}
    prediction\_consistency=\frac{\sum_{i=1}^N \mathcal{F}(x_i)==R(x_i)}{N}
\end{equation}

Table~\ref{tab:faithfulness} shows the prediction consistency of the two models in each task on the training and test data separately. We can see that for the binary classification models (Toxic and Quora), the abstraction models can provide highly consistent predictions (consistency is 99\% and 97\%) compared with the original RNN models on both training and test data. For the multi-classification model (AGNews), the abstraction model can still provide very consistent predictions (consistency is 86\%). These results demonstrate the faithfulness of our state abstraction technique.

\section{Number of Abstracted States}
\label{appendix:states}

\begin{figure}[t]
    \centering
    \begin{subfigure}[b]{0.42\textwidth}
         \centering
         \includegraphics[width=\textwidth]{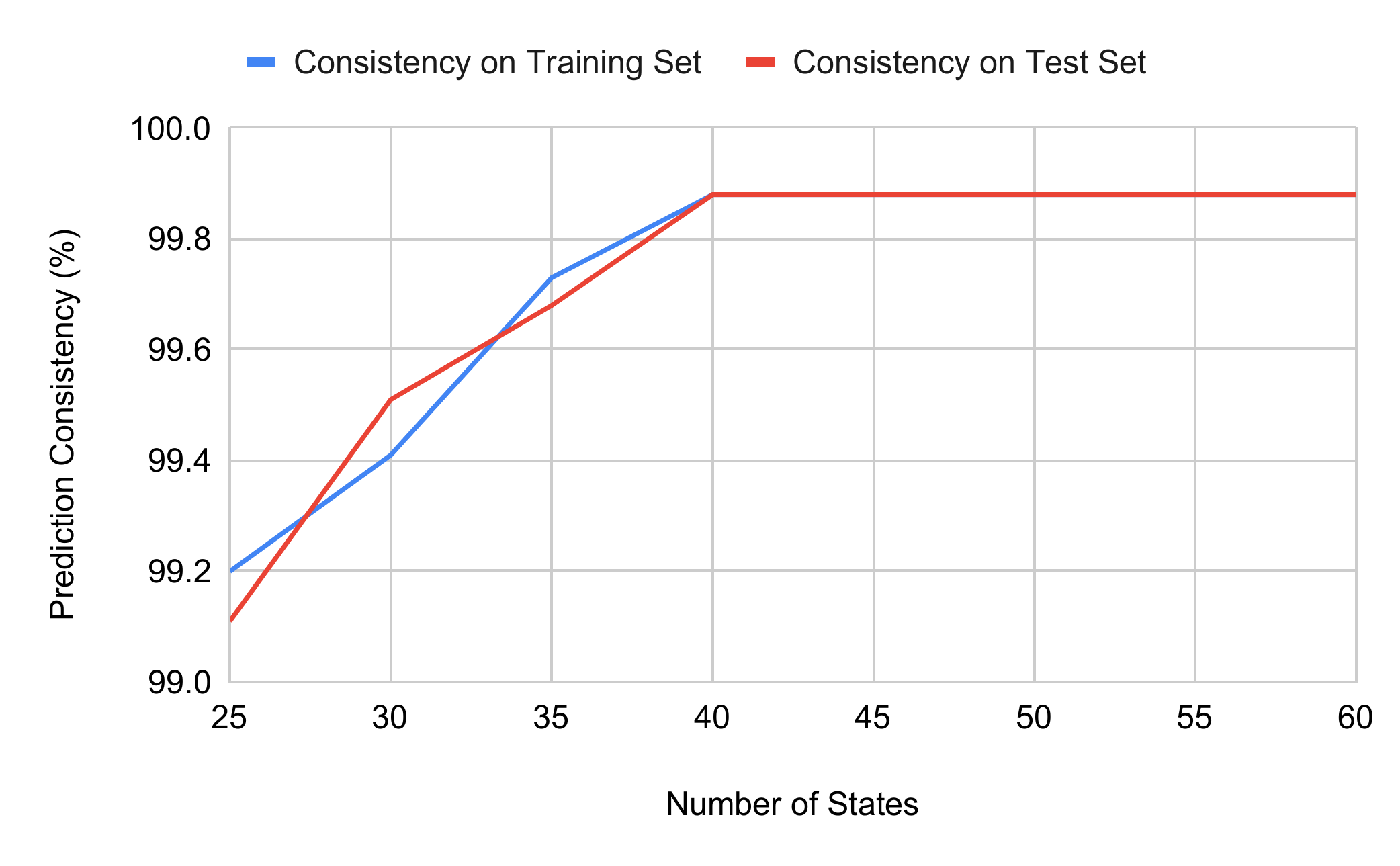}
         \caption{\responseref{} Prediction consistency between the original RNN model and the abstracted model on Toxic dataset.}
         \label{fig:num_states_toxic}
    \end{subfigure}
    \vfill
    \begin{subfigure}[b]{0.42\textwidth}
         \centering
         \includegraphics[width=\textwidth]{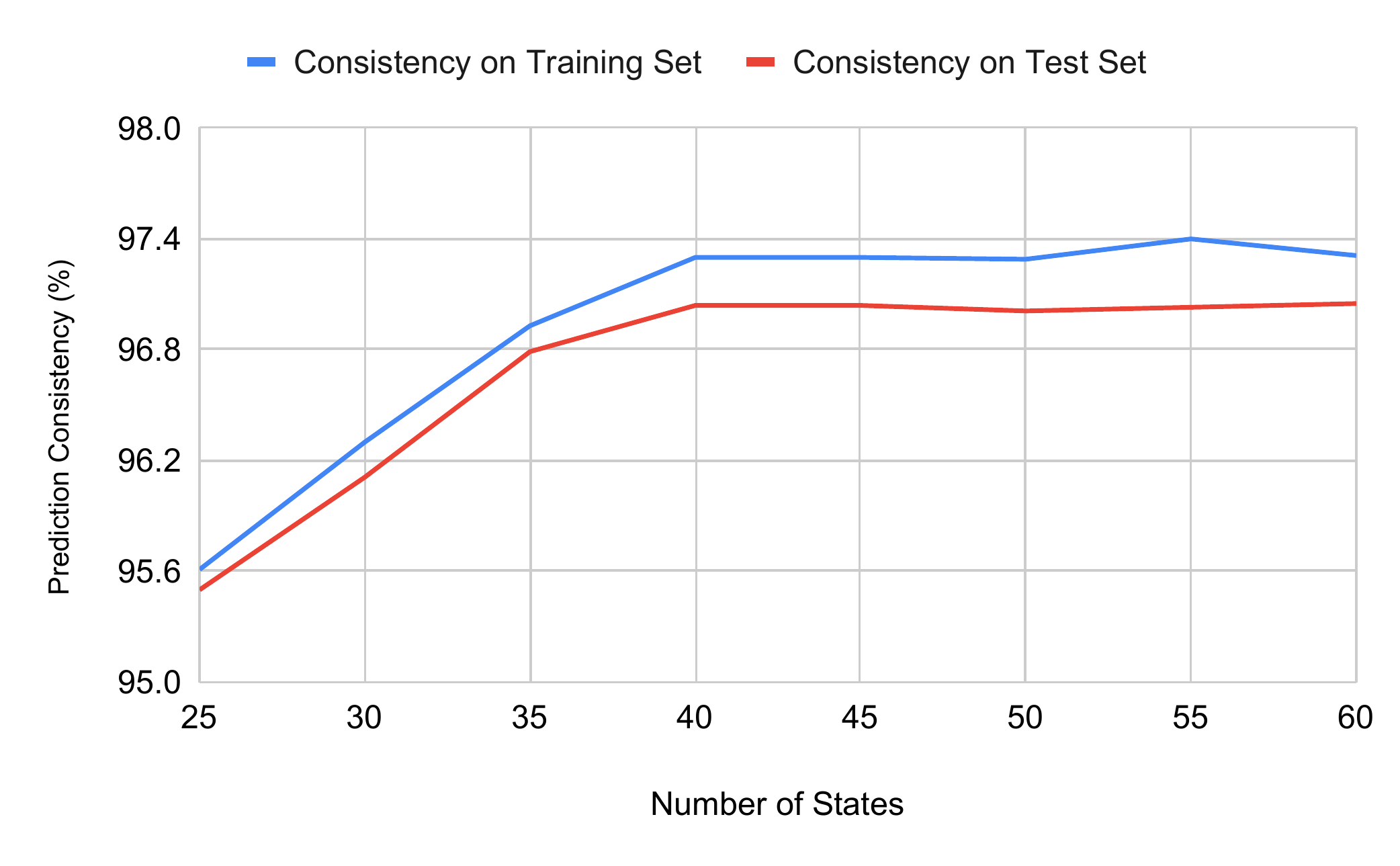}
         \caption{\responseref{} Prediction consistency between the original RNN model and the abstracted model on Quora dataset.}
         \label{fig:num_states_quora}
    \end{subfigure}
    \vfill
    \begin{subfigure}[b]{0.42\textwidth}
         \centering
         \includegraphics[width=\textwidth]{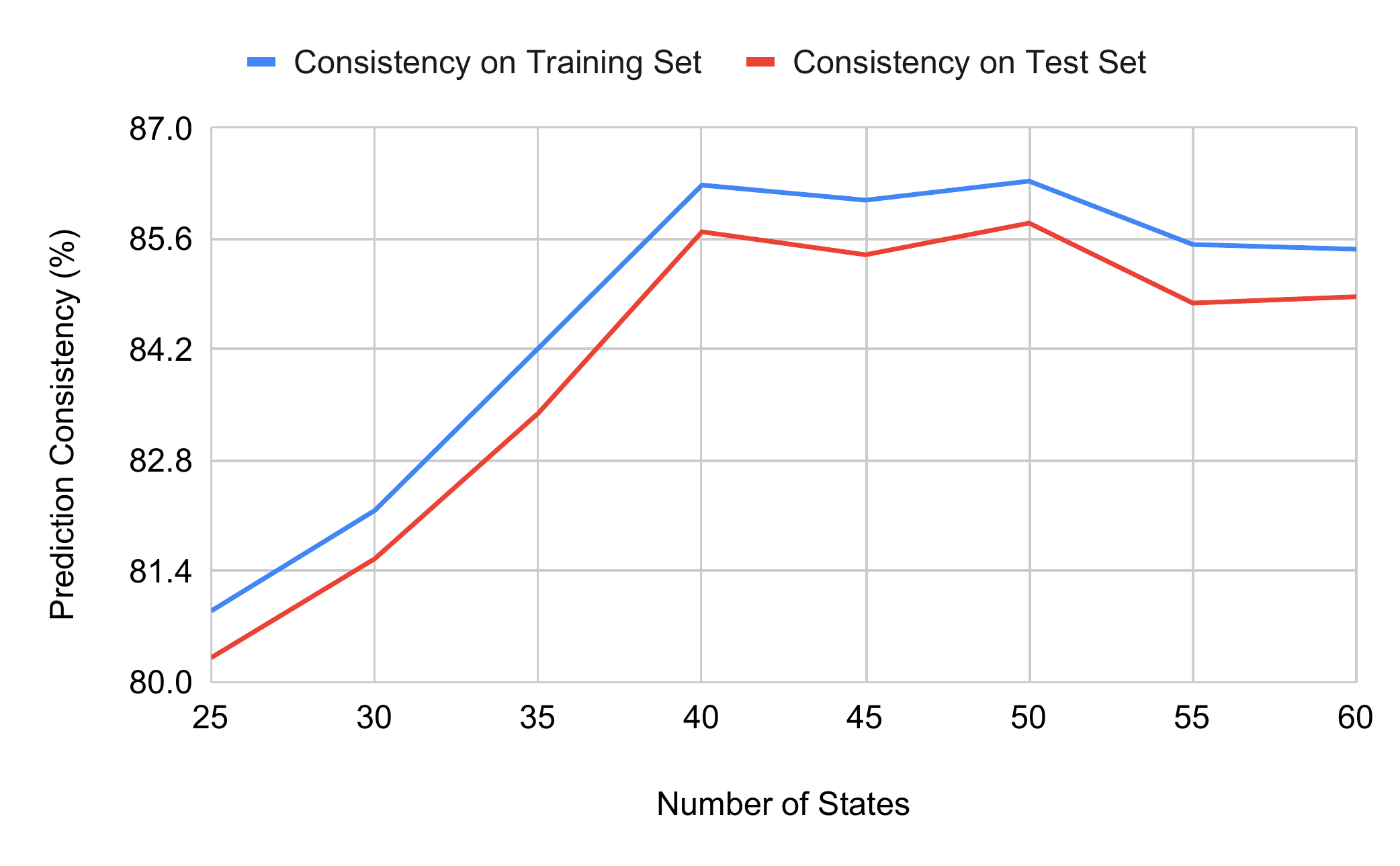}
         \caption{\responseref{} Prediction consistency between the original RNN model and the abstracted model on AGNews dataset.}
         \label{fig:num_states_agnews}
    \end{subfigure}
    \caption{\responseref{} Prediction consistency w.r.t. the number of states of all three RNN models' abstracted models.}
    \Description{This figure shows the prediction consistency w.r.t the different number of states on three different tasks (Toxic, Quora, and AGNews). It can be seen that when the number of states is smaller than 40, the prediction consistency is low. However, when the number of states is larger than 40, there is not much improvement in prediction consistency.}
    \label{fig:num_states}
\end{figure}

During our user study, we set the number of abstracted states to 40. This number is empirically decided to achieve a good balance between the prediction consistency to the original RNN and the cognitive effort of inspecting a state diagram. To further show that this setting will not affect the abstraction model's faithfulness, we report the abstracted models' prediction consistency w.r.t. the number of states of all three models in Fig.~\ref{fig:num_states}.

As we can see, a lower number of states will lead to a lower prediction consistency. However, when the number of states is larger than 40, the prediction consistency stays largely the same. Therefore, we choose this number of states, 40, throughout our motivating example and user study.

}

\section{ML Tasks}
\label{appendix:interface}

\subsection{ML Task 1: Sentimental Analysis (Quora dataset)}

{\bf Task Description:}

In this task, participants were given an RNN model trained on the Quora dataset. 

Quora dataset is collected from quora.com, where each text in the dataset is labeled as ``Sincere'' or ``Insincere''. An insincere question is defined as a question intended to make a statement rather than look for helpful answers.

Participants first used the tool to understand the model. They were asked to use the tool to explore the model’s behaviours and performance on training data and test data. After exploring, participants were asked to share their findings, e.g., Did they find any insights? Did they find any bugs? How would they improve this model?

Then participants were given 5 misclassified sentences. Participants had 10 minutes in total to finish the following task: for each sentence, participants needed to find out why this sentence is misclassified with the help of DeepSeer. 

\begin{figure*}[t]
\centering
  \includegraphics[width=\linewidth]{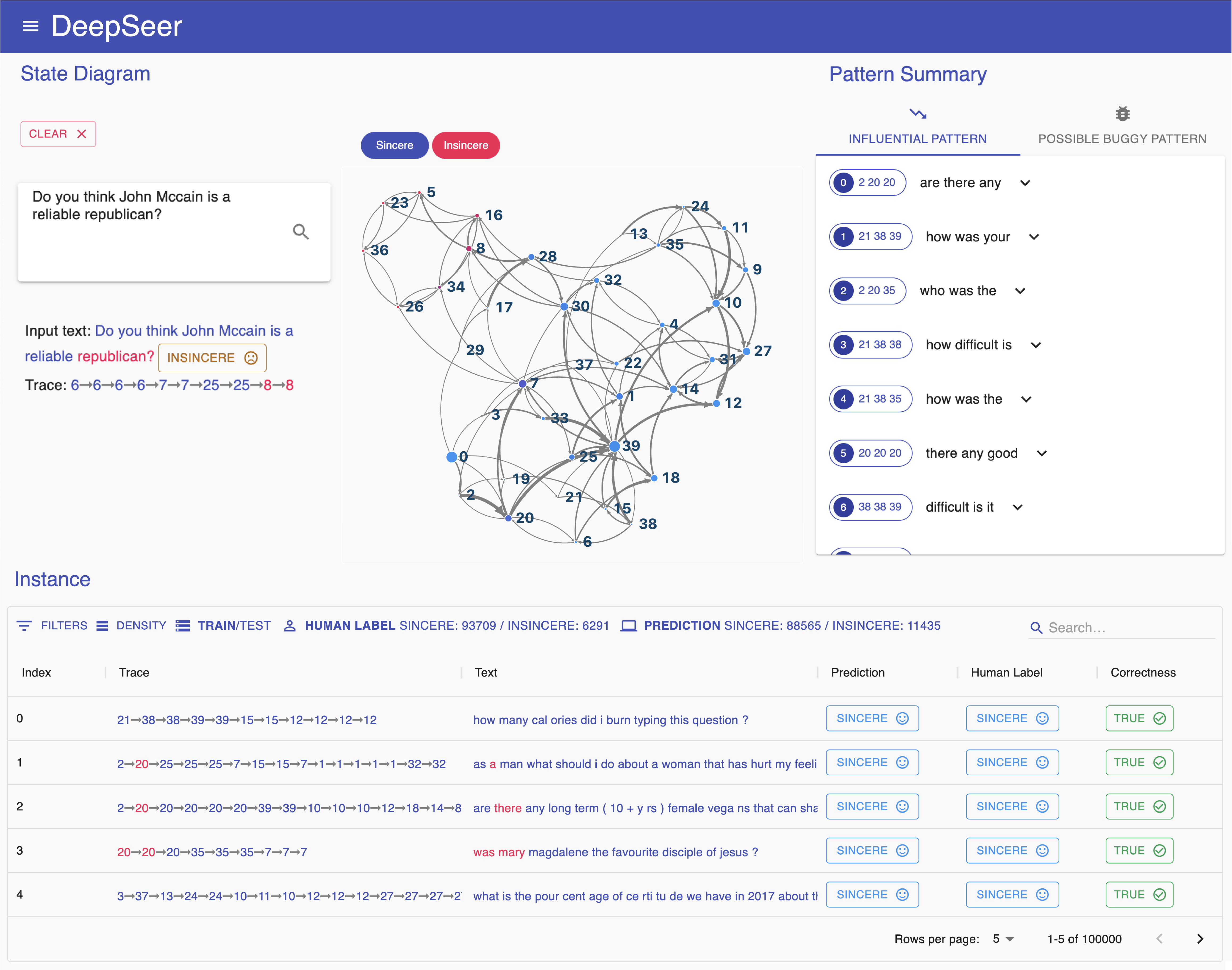}
  \caption{The interface of {\tool} used for ML task 1 (Quora dataset).}
  \Description{This figure displays the interface of DeepSeer on Quora dataset.}
  \label{fig:quora}
\end{figure*}

\subsection{ML Task 2: Topic Classification (AGNews dataset)}

{\bf Task Description:}

In this task, participants were given an RNN model trained on the AGNews dataset.

AGNews dataset is a collection of news articles. This RNN model classifies each text into different topics, including World, Sports, Business, and Sci/Tech. 

Participants first used the tool to understand the model. They were asked to use the tool to explore the model’s behaviours, and performance on training data and test data. After exploring, participants were asked to share their findings, e.g., Did they find any insights? Did they find any bugs? How would they improve this model?

Then participants were given 5 misclassified sentences. Participants had 10 minutes in total to finish the following task: for each sentence, participants needed to find out why this sentence is misclassified with the help of DeepSeer. 

\begin{figure*}[t]
\centering
  \includegraphics[width=\linewidth]{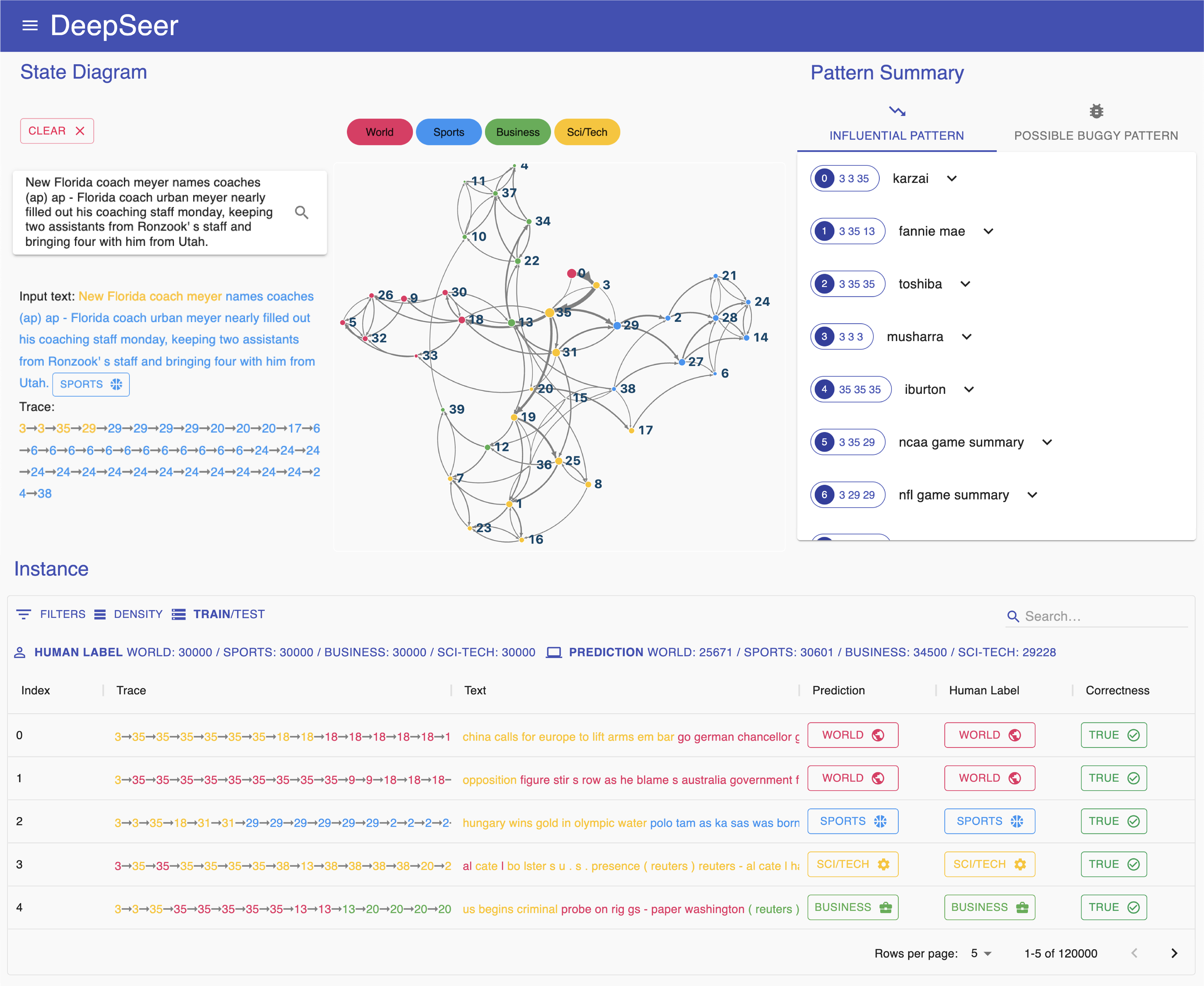}
  \caption{The interface of {\tool} used for ML task 2 (AGNews dataset).}
  \Description{This figure displays the interface of DeepSeer on AGNews dataset.}
  \label{fig:agnews}
\end{figure*}

%% file: main.bbl

\begin{thebibliography}{65}

%%% the Web address.

\ifx \showCODEN    \undefined \def \showCODEN     #1{\unskip}     \fi
\ifx \showDOI      \undefined \def \showDOI       #1{#1}\fi
\ifx \showISBNx    \undefined \def \showISBNx     #1{\unskip}     \fi
\ifx \showISBNxiii \undefined \def \showISBNxiii  #1{\unskip}     \fi
\ifx \showISSN     \undefined \def \showISSN      #1{\unskip}     \fi
\ifx \showLCCN     \undefined \def \showLCCN      #1{\unskip}     \fi
\ifx \shownote     \undefined \def \shownote      #1{#1}          \fi
\ifx \showarticletitle \undefined \def \showarticletitle #1{#1}   \fi
\ifx \showURL      \undefined \def \showURL       {\relax}        \fi
% invisible to TeX
\providecommand\bibfield[2]{#2}
\providecommand\bibinfo[2]{#2}
\providecommand\natexlab[1]{#1}
\providecommand\showeprint[2][]{arXiv:#2}

\bibitem[Abadi et~al\mbox{.}(2015)]%
        {tensorflow2015-whitepaper}
\bibfield{author}{\bibinfo{person}{Mart\'{\i}n Abadi}, \bibinfo{person}{Ashish
  Agarwal}, \bibinfo{person}{Paul Barham}, \bibinfo{person}{Eugene Brevdo},
  \bibinfo{person}{Zhifeng Chen}, \bibinfo{person}{Craig Citro},
  \bibinfo{person}{Greg~S. Corrado}, \bibinfo{person}{Andy Davis},
  \bibinfo{person}{Jeffrey Dean}, \bibinfo{person}{Matthieu Devin},
  \bibinfo{person}{Sanjay Ghemawat}, \bibinfo{person}{Ian Goodfellow},
  \bibinfo{person}{Andrew Harp}, \bibinfo{person}{Geoffrey Irving},
  \bibinfo{person}{Michael Isard}, \bibinfo{person}{Yangqing Jia},
  \bibinfo{person}{Rafal Jozefowicz}, \bibinfo{person}{Lukasz Kaiser},
  \bibinfo{person}{Manjunath Kudlur}, \bibinfo{person}{Josh Levenberg},
  \bibinfo{person}{Dandelion Man\'{e}}, \bibinfo{person}{Rajat Monga},
  \bibinfo{person}{Sherry Moore}, \bibinfo{person}{Derek Murray},
  \bibinfo{person}{Chris Olah}, \bibinfo{person}{Mike Schuster},
  \bibinfo{person}{Jonathon Shlens}, \bibinfo{person}{Benoit Steiner},
  \bibinfo{person}{Ilya Sutskever}, \bibinfo{person}{Kunal Talwar},
  \bibinfo{person}{Paul Tucker}, \bibinfo{person}{Vincent Vanhoucke},
  \bibinfo{person}{Vijay Vasudevan}, \bibinfo{person}{Fernanda Vi\'{e}gas},
  \bibinfo{person}{Oriol Vinyals}, \bibinfo{person}{Pete Warden},
  \bibinfo{person}{Martin Wattenberg}, \bibinfo{person}{Martin Wicke},
  \bibinfo{person}{Yuan Yu}, {and} \bibinfo{person}{Xiaoqiang Zheng}.}
  \bibinfo{year}{2015}\natexlab{}.
\newblock \bibinfo{title}{{TensorFlow}: Large-Scale Machine Learning on
  Heterogeneous Systems}.
\newblock
\newblock
\urldef\tempurl%
\url{https://www.tensorflow.org/}
\showURL{%
\tempurl}
\newblock
\shownote{Software available from tensorflow.org}.


\bibitem[Adadi and Berrada(2018)]%
        {adadi2018peeking}
\bibfield{author}{\bibinfo{person}{Amina Adadi} {and} \bibinfo{person}{Mohammed
  Berrada}.} \bibinfo{year}{2018}\natexlab{}.
\newblock \showarticletitle{Peeking inside the black-box: a survey on
  explainable artificial intelligence (XAI)}.
\newblock \bibinfo{journal}{\emph{IEEE access}}  \bibinfo{volume}{6}
  (\bibinfo{year}{2018}), \bibinfo{pages}{52138--52160}.
\newblock


\bibitem[Adebayo et~al\mbox{.}(2020)]%
        {adebayo2020debugging}
\bibfield{author}{\bibinfo{person}{Julius Adebayo}, \bibinfo{person}{Michael
  Muelly}, \bibinfo{person}{Ilaria Liccardi}, {and} \bibinfo{person}{Been
  Kim}.} \bibinfo{year}{2020}\natexlab{}.
\newblock \showarticletitle{Debugging Tests for Model Explanations}.
\newblock \bibinfo{journal}{\emph{Advances in Neural Information Processing
  Systems}}  \bibinfo{volume}{33} (\bibinfo{year}{2020}),
  \bibinfo{pages}{700--712}.
\newblock


\bibitem[Amershi et~al\mbox{.}(2015)]%
        {amershi2015modeltracker}
\bibfield{author}{\bibinfo{person}{Saleema Amershi}, \bibinfo{person}{Max
  Chickering}, \bibinfo{person}{Steven~M Drucker}, \bibinfo{person}{Bongshin
  Lee}, \bibinfo{person}{Patrice Simard}, {and} \bibinfo{person}{Jina Suh}.}
  \bibinfo{year}{2015}\natexlab{}.
\newblock \showarticletitle{Modeltracker: Redesigning performance analysis
  tools for machine learning}. In \bibinfo{booktitle}{\emph{Proceedings of the
  33rd Annual ACM Conference on Human Factors in Computing Systems}}.
  \bibinfo{pages}{337--346}.
\newblock


\bibitem[Amershi et~al\mbox{.}(2019)]%
        {amershi2019guidelines}
\bibfield{author}{\bibinfo{person}{Saleema Amershi}, \bibinfo{person}{Dan
  Weld}, \bibinfo{person}{Mihaela Vorvoreanu}, \bibinfo{person}{Adam Fourney},
  \bibinfo{person}{Besmira Nushi}, \bibinfo{person}{Penny Collisson},
  \bibinfo{person}{Jina Suh}, \bibinfo{person}{Shamsi Iqbal},
  \bibinfo{person}{Paul~N Bennett}, \bibinfo{person}{Kori Inkpen},
  {et~al\mbox{.}}} \bibinfo{year}{2019}\natexlab{}.
\newblock \showarticletitle{Guidelines for human-AI interaction}. In
  \bibinfo{booktitle}{\emph{Proceedings of the 2019 chi conference on human
  factors in computing systems}}. \bibinfo{pages}{1--13}.
\newblock


\bibitem[Arrieta et~al\mbox{.}(2020)]%
        {arrieta2020explainable}
\bibfield{author}{\bibinfo{person}{Alejandro~Barredo Arrieta},
  \bibinfo{person}{Natalia D{\'\i}az-Rodr{\'\i}guez}, \bibinfo{person}{Javier
  Del~Ser}, \bibinfo{person}{Adrien Bennetot}, \bibinfo{person}{Siham Tabik},
  \bibinfo{person}{Alberto Barbado}, \bibinfo{person}{Salvador Garc{\'\i}a},
  \bibinfo{person}{Sergio Gil-L{\'o}pez}, \bibinfo{person}{Daniel Molina},
  \bibinfo{person}{Richard Benjamins}, {et~al\mbox{.}}}
  \bibinfo{year}{2020}\natexlab{}.
\newblock \showarticletitle{Explainable Artificial Intelligence (XAI):
  Concepts, taxonomies, opportunities and challenges toward responsible AI}.
\newblock \bibinfo{journal}{\emph{Information Fusion}}  \bibinfo{volume}{58}
  (\bibinfo{year}{2020}), \bibinfo{pages}{82--115}.
\newblock


\bibitem[Barshan et~al\mbox{.}(2020)]%
        {barshan2020relatif}
\bibfield{author}{\bibinfo{person}{Elnaz Barshan},
  \bibinfo{person}{Marc-Etienne Brunet}, {and}
  \bibinfo{person}{Gintare~Karolina Dziugaite}.}
  \bibinfo{year}{2020}\natexlab{}.
\newblock \showarticletitle{Relatif: Identifying explanatory training samples
  via relative influence}. In \bibinfo{booktitle}{\emph{International
  Conference on Artificial Intelligence and Statistics}}. PMLR,
  \bibinfo{pages}{1899--1909}.
\newblock


\bibitem[Bau et~al\mbox{.}(2017)]%
        {bau2017network}
\bibfield{author}{\bibinfo{person}{David Bau}, \bibinfo{person}{Bolei Zhou},
  \bibinfo{person}{Aditya Khosla}, \bibinfo{person}{Aude Oliva}, {and}
  \bibinfo{person}{Antonio Torralba}.} \bibinfo{year}{2017}\natexlab{}.
\newblock \showarticletitle{Network dissection: Quantifying interpretability of
  deep visual representations}. In \bibinfo{booktitle}{\emph{Proceedings of the
  IEEE conference on computer vision and pattern recognition}}.
  \bibinfo{pages}{6541--6549}.
\newblock


\bibitem[Bocklisch et~al\mbox{.}(2017)]%
        {bocklisch2017rasa}
\bibfield{author}{\bibinfo{person}{Tom Bocklisch}, \bibinfo{person}{Joey
  Faulkner}, \bibinfo{person}{Nick Pawlowski}, {and} \bibinfo{person}{Alan
  Nichol}.} \bibinfo{year}{2017}\natexlab{}.
\newblock \showarticletitle{Rasa: Open source language understanding and
  dialogue management}.
\newblock \bibinfo{journal}{\emph{arXiv preprint arXiv:1712.05181}}
  (\bibinfo{year}{2017}).
\newblock


\bibitem[B{\"o}gl et~al\mbox{.}(2013)]%
        {bogl2013visual}
\bibfield{author}{\bibinfo{person}{Markus B{\"o}gl}, \bibinfo{person}{Wolfgang
  Aigner}, \bibinfo{person}{Peter Filzmoser}, \bibinfo{person}{Tim Lammarsch},
  \bibinfo{person}{Silvia Miksch}, {and} \bibinfo{person}{Alexander Rind}.}
  \bibinfo{year}{2013}\natexlab{}.
\newblock \showarticletitle{Visual analytics for model selection in time series
  analysis}.
\newblock \bibinfo{journal}{\emph{IEEE transactions on visualization and
  computer graphics}} \bibinfo{volume}{19}, \bibinfo{number}{12}
  (\bibinfo{year}{2013}), \bibinfo{pages}{2237--2246}.
\newblock


\bibitem[Cai et~al\mbox{.}(2019)]%
        {cai2019effects}
\bibfield{author}{\bibinfo{person}{Carrie~J Cai}, \bibinfo{person}{Jonas
  Jongejan}, {and} \bibinfo{person}{Jess Holbrook}.}
  \bibinfo{year}{2019}\natexlab{}.
\newblock \showarticletitle{The effects of example-based explanations in a
  machine learning interface}. In \bibinfo{booktitle}{\emph{Proceedings of the
  24th international conference on intelligent user interfaces}}.
  \bibinfo{pages}{258--262}.
\newblock


\bibitem[Cho et~al\mbox{.}(2014)]%
        {cho2014properties}
\bibfield{author}{\bibinfo{person}{Kyunghyun Cho}, \bibinfo{person}{Bart van
  Merri{\"e}nboer}, \bibinfo{person}{Dzmitry Bahdanau}, {and}
  \bibinfo{person}{Yoshua Bengio}.} \bibinfo{year}{2014}\natexlab{}.
\newblock \showarticletitle{On the Properties of Neural Machine Translation:
  Encoder--Decoder Approaches}. In \bibinfo{booktitle}{\emph{Proceedings of
  SSST-8, Eighth Workshop on Syntax, Semantics and Structure in Statistical
  Translation}}. \bibinfo{pages}{103--111}.
\newblock


\bibitem[Das and Rad(2020)]%
        {das2020opportunities}
\bibfield{author}{\bibinfo{person}{Arun Das} {and} \bibinfo{person}{Paul Rad}.}
  \bibinfo{year}{2020}\natexlab{}.
\newblock \showarticletitle{Opportunities and challenges in explainable
  artificial intelligence (xai): A survey}.
\newblock \bibinfo{journal}{\emph{arXiv preprint arXiv:2006.11371}}
  (\bibinfo{year}{2020}).
\newblock


\bibitem[Dodge et~al\mbox{.}(2019)]%
        {dodge2019explaining}
\bibfield{author}{\bibinfo{person}{Jonathan Dodge}, \bibinfo{person}{Q~Vera
  Liao}, \bibinfo{person}{Yunfeng Zhang}, \bibinfo{person}{Rachel~KE Bellamy},
  {and} \bibinfo{person}{Casey Dugan}.} \bibinfo{year}{2019}\natexlab{}.
\newblock \showarticletitle{Explaining models: an empirical study of how
  explanations impact fairness judgment}. In
  \bibinfo{booktitle}{\emph{Proceedings of the 24th international conference on
  intelligent user interfaces}}. \bibinfo{pages}{275--285}.
\newblock


\bibitem[Du et~al\mbox{.}(2020)]%
        {du2020marble}
\bibfield{author}{\bibinfo{person}{Xiaoning Du}, \bibinfo{person}{Yi Li},
  \bibinfo{person}{Xiaofei Xie}, \bibinfo{person}{Lei Ma},
  \bibinfo{person}{Yang Liu}, {and} \bibinfo{person}{Jianjun Zhao}.}
  \bibinfo{year}{2020}\natexlab{}.
\newblock \showarticletitle{Marble: Model-based robustness analysis of stateful
  deep learning systems}. In \bibinfo{booktitle}{\emph{Proceedings of the 35th
  IEEE/ACM International Conference on Automated Software Engineering}}.
  \bibinfo{pages}{423--435}.
\newblock


\bibitem[Du et~al\mbox{.}(2019)]%
        {du2019deepstellar}
\bibfield{author}{\bibinfo{person}{Xiaoning Du}, \bibinfo{person}{Xiaofei Xie},
  \bibinfo{person}{Yi Li}, \bibinfo{person}{Lei Ma}, \bibinfo{person}{Yang
  Liu}, {and} \bibinfo{person}{Jianjun Zhao}.} \bibinfo{year}{2019}\natexlab{}.
\newblock \showarticletitle{Deepstellar: Model-based quantitative analysis of
  stateful deep learning systems}. In \bibinfo{booktitle}{\emph{Proceedings of
  the 2019 27th ACM Joint Meeting on European Software Engineering Conference
  and Symposium on the Foundations of Software Engineering}}.
  \bibinfo{pages}{477--487}.
\newblock


\bibitem[Dzindolet et~al\mbox{.}(2003)]%
        {dzindolet2003role}
\bibfield{author}{\bibinfo{person}{Mary~T Dzindolet}, \bibinfo{person}{Scott~A
  Peterson}, \bibinfo{person}{Regina~A Pomranky}, \bibinfo{person}{Linda~G
  Pierce}, {and} \bibinfo{person}{Hall~P Beck}.}
  \bibinfo{year}{2003}\natexlab{}.
\newblock \showarticletitle{The role of trust in automation reliance}.
\newblock \bibinfo{journal}{\emph{International journal of human-computer
  studies}} \bibinfo{volume}{58}, \bibinfo{number}{6} (\bibinfo{year}{2003}),
  \bibinfo{pages}{697--718}.
\newblock


\bibitem[Elman(1990)]%
        {elman1990finding}
\bibfield{author}{\bibinfo{person}{Jeffrey~L Elman}.}
  \bibinfo{year}{1990}\natexlab{}.
\newblock \showarticletitle{Finding structure in time}.
\newblock \bibinfo{journal}{\emph{Cognitive science}} \bibinfo{volume}{14},
  \bibinfo{number}{2} (\bibinfo{year}{1990}), \bibinfo{pages}{179--211}.
\newblock


\bibitem[Fournier-Viger et~al\mbox{.}(2013)]%
        {fournier2013tks}
\bibfield{author}{\bibinfo{person}{Philippe Fournier-Viger},
  \bibinfo{person}{Antonio Gomariz}, \bibinfo{person}{Ted Gueniche},
  \bibinfo{person}{Esp{\'e}rance Mwamikazi}, {and} \bibinfo{person}{Rincy
  Thomas}.} \bibinfo{year}{2013}\natexlab{}.
\newblock \showarticletitle{TKS: efficient mining of top-k sequential
  patterns}. In \bibinfo{booktitle}{\emph{International Conference on Advanced
  Data Mining and Applications}}. Springer, \bibinfo{pages}{109--120}.
\newblock


\bibitem[Gulli(2005)]%
        {agnews}
\bibfield{author}{\bibinfo{person}{Antonio Gulli}.}
  \bibinfo{year}{2005}\natexlab{}.
\newblock \bibinfo{title}{AG's corpus of news articles}.
\newblock
\newblock
\urldef\tempurl%
\url{http://groups.di.unipi.it/~gulli/AG_corpus_of_news_articles.html}
\showURL{%
\tempurl}


\bibitem[Hart and Staveland(1988)]%
        {hart1988development}
\bibfield{author}{\bibinfo{person}{Sandra~G Hart} {and}
  \bibinfo{person}{Lowell~E Staveland}.} \bibinfo{year}{1988}\natexlab{}.
\newblock \showarticletitle{Development of NASA-TLX (Task Load Index): Results
  of empirical and theoretical research}.
\newblock In \bibinfo{booktitle}{\emph{Advances in psychology}}.
  Vol.~\bibinfo{volume}{52}. \bibinfo{publisher}{Elsevier},
  \bibinfo{pages}{139--183}.
\newblock


\bibitem[Herlocker et~al\mbox{.}(2000)]%
        {herlocker2000explaining}
\bibfield{author}{\bibinfo{person}{Jonathan~L Herlocker},
  \bibinfo{person}{Joseph~A Konstan}, {and} \bibinfo{person}{John Riedl}.}
  \bibinfo{year}{2000}\natexlab{}.
\newblock \showarticletitle{Explaining collaborative filtering
  recommendations}. In \bibinfo{booktitle}{\emph{Proceedings of the 2000 ACM
  conference on Computer supported cooperative work}}.
  \bibinfo{pages}{241--250}.
\newblock


\bibitem[Hochreiter and Schmidhuber(1997)]%
        {hochreiter1997long}
\bibfield{author}{\bibinfo{person}{Sepp Hochreiter} {and}
  \bibinfo{person}{J{\"u}rgen Schmidhuber}.} \bibinfo{year}{1997}\natexlab{}.
\newblock \showarticletitle{Long short-term memory}.
\newblock \bibinfo{journal}{\emph{Neural computation}} \bibinfo{volume}{9},
  \bibinfo{number}{8} (\bibinfo{year}{1997}), \bibinfo{pages}{1735--1780}.
\newblock


\bibitem[Hohman et~al\mbox{.}(2019)]%
        {hohman2019gamut}
\bibfield{author}{\bibinfo{person}{Fred Hohman}, \bibinfo{person}{Andrew Head},
  \bibinfo{person}{Rich Caruana}, \bibinfo{person}{Robert DeLine}, {and}
  \bibinfo{person}{Steven~M Drucker}.} \bibinfo{year}{2019}\natexlab{}.
\newblock \showarticletitle{Gamut: A design probe to understand how data
  scientists understand machine learning models}. In
  \bibinfo{booktitle}{\emph{Proceedings of the 2019 CHI conference on human
  factors in computing systems}}. \bibinfo{pages}{1--13}.
\newblock


\bibitem[Jin et~al\mbox{.}(2022)]%
        {jin2022gnnlens}
\bibfield{author}{\bibinfo{person}{Zhihua Jin}, \bibinfo{person}{Yong Wang},
  \bibinfo{person}{Qianwen Wang}, \bibinfo{person}{Yao Ming},
  \bibinfo{person}{Tengfei Ma}, {and} \bibinfo{person}{Huamin Qu}.}
  \bibinfo{year}{2022}\natexlab{}.
\newblock \showarticletitle{Gnnlens: A visual analytics approach for prediction
  error diagnosis of graph neural networks}.
\newblock \bibinfo{journal}{\emph{IEEE Transactions on Visualization and
  Computer Graphics}} (\bibinfo{year}{2022}).
\newblock


\bibitem[Kahng et~al\mbox{.}(2017)]%
        {kahng2017cti}
\bibfield{author}{\bibinfo{person}{Minsuk Kahng}, \bibinfo{person}{Pierre~Y
  Andrews}, \bibinfo{person}{Aditya Kalro}, {and} \bibinfo{person}{Duen~Horng
  Chau}.} \bibinfo{year}{2017}\natexlab{}.
\newblock \showarticletitle{Activis: Visual exploration of industry-scale deep
  neural network models}.
\newblock \bibinfo{journal}{\emph{IEEE transactions on visualization and
  computer graphics}} \bibinfo{volume}{24}, \bibinfo{number}{1}
  (\bibinfo{year}{2017}), \bibinfo{pages}{88--97}.
\newblock


\bibitem[Karpathy et~al\mbox{.}(2015)]%
        {karpathy2015visualizing}
\bibfield{author}{\bibinfo{person}{Andrej Karpathy}, \bibinfo{person}{Justin
  Johnson}, {and} \bibinfo{person}{Li Fei-Fei}.}
  \bibinfo{year}{2015}\natexlab{}.
\newblock \showarticletitle{Visualizing and understanding recurrent networks}.
\newblock \bibinfo{journal}{\emph{arXiv preprint arXiv:1506.02078}}
  (\bibinfo{year}{2015}).
\newblock


\bibitem[Kaur et~al\mbox{.}(2020)]%
        {kaur2020interpreting}
\bibfield{author}{\bibinfo{person}{Harmanpreet Kaur}, \bibinfo{person}{Harsha
  Nori}, \bibinfo{person}{Samuel Jenkins}, \bibinfo{person}{Rich Caruana},
  \bibinfo{person}{Hanna Wallach}, {and} \bibinfo{person}{Jennifer
  Wortman~Vaughan}.} \bibinfo{year}{2020}\natexlab{}.
\newblock \showarticletitle{Interpreting Interpretability: Understanding Data
  Scientists' Use of Interpretability Tools for Machine Learning}. In
  \bibinfo{booktitle}{\emph{Proceedings of the 2020 CHI Conference on Human
  Factors in Computing Systems}}. \bibinfo{pages}{1--14}.
\newblock


\bibitem[Kim et~al\mbox{.}(2018)]%
        {kim2018interpretability}
\bibfield{author}{\bibinfo{person}{Been Kim}, \bibinfo{person}{Martin
  Wattenberg}, \bibinfo{person}{Justin Gilmer}, \bibinfo{person}{Carrie Cai},
  \bibinfo{person}{James Wexler}, \bibinfo{person}{Fernanda Viegas},
  {et~al\mbox{.}}} \bibinfo{year}{2018}\natexlab{}.
\newblock \showarticletitle{Interpretability beyond feature attribution:
  Quantitative testing with concept activation vectors (tcav)}. In
  \bibinfo{booktitle}{\emph{International conference on machine learning}}.
  PMLR, \bibinfo{pages}{2668--2677}.
\newblock


\bibitem[Kizilcec(2016)]%
        {kizilcec2016much}
\bibfield{author}{\bibinfo{person}{Ren{\'e}~F Kizilcec}.}
  \bibinfo{year}{2016}\natexlab{}.
\newblock \showarticletitle{How much information? Effects of transparency on
  trust in an algorithmic interface}. In \bibinfo{booktitle}{\emph{Proceedings
  of the 2016 CHI Conference on Human Factors in Computing Systems}}.
  \bibinfo{pages}{2390--2395}.
\newblock


\bibitem[Koh et~al\mbox{.}(2019)]%
        {KohATL19}
\bibfield{author}{\bibinfo{person}{Pang~Wei Koh}, \bibinfo{person}{Kai{-}Siang
  Ang}, \bibinfo{person}{Hubert H.~K. Teo}, {and} \bibinfo{person}{Percy
  Liang}.} \bibinfo{year}{2019}\natexlab{}.
\newblock \showarticletitle{On the Accuracy of Influence Functions for
  Measuring Group Effects}. In \bibinfo{booktitle}{\emph{Advances in Neural
  Information Processing Systems 32 (2019)}}. \bibinfo{pages}{5255--5265}.
\newblock


\bibitem[Koh and Liang(2017)]%
        {koh2017understanding}
\bibfield{author}{\bibinfo{person}{Pang~Wei Koh} {and} \bibinfo{person}{Percy
  Liang}.} \bibinfo{year}{2017}\natexlab{}.
\newblock \showarticletitle{Understanding black-box predictions via influence
  functions}. In \bibinfo{booktitle}{\emph{International Conference on Machine
  Learning}}. PMLR, \bibinfo{pages}{1885--1894}.
\newblock


\bibitem[Li et~al\mbox{.}(2016)]%
        {li2015visualizing}
\bibfield{author}{\bibinfo{person}{Jiwei Li}, \bibinfo{person}{Xinlei Chen},
  \bibinfo{person}{Eduard Hovy}, {and} \bibinfo{person}{Dan Jurafsky}.}
  \bibinfo{year}{2016}\natexlab{}.
\newblock \showarticletitle{Visualizing and Understanding Neural Models in
  NLP}. In \bibinfo{booktitle}{\emph{Proceedings of the 2016 Conference of the
  North American Chapter of the Association for Computational Linguistics:
  Human Language Technologies}}. \bibinfo{pages}{681--691}.
\newblock


\bibitem[Li et~al\mbox{.}(2014)]%
        {li2014deep}
\bibfield{author}{\bibinfo{person}{Rongjian Li}, \bibinfo{person}{Wenlu Zhang},
  \bibinfo{person}{Heung-Il Suk}, \bibinfo{person}{Li Wang},
  \bibinfo{person}{Jiang Li}, \bibinfo{person}{Dinggang Shen}, {and}
  \bibinfo{person}{Shuiwang Ji}.} \bibinfo{year}{2014}\natexlab{}.
\newblock \showarticletitle{Deep learning based imaging data completion for
  improved brain disease diagnosis}. In \bibinfo{booktitle}{\emph{International
  conference on medical image computing and computer-assisted intervention}}.
  Springer, \bibinfo{pages}{305--312}.
\newblock


\bibitem[Liao et~al\mbox{.}(2020)]%
        {liao2020questioning}
\bibfield{author}{\bibinfo{person}{Q~Vera Liao}, \bibinfo{person}{Daniel
  Gruen}, {and} \bibinfo{person}{Sarah Miller}.}
  \bibinfo{year}{2020}\natexlab{}.
\newblock \showarticletitle{Questioning the AI: informing design practices for
  explainable AI user experiences}. In \bibinfo{booktitle}{\emph{Proceedings of
  the 2020 CHI Conference on Human Factors in Computing Systems}}.
  \bibinfo{pages}{1--15}.
\newblock


\bibitem[Lipton(2018)]%
        {lipton2018mythos}
\bibfield{author}{\bibinfo{person}{Zachary~C Lipton}.}
  \bibinfo{year}{2018}\natexlab{}.
\newblock \showarticletitle{The Mythos of Model Interpretability: In machine
  learning, the concept of interpretability is both important and slippery.}
\newblock \bibinfo{journal}{\emph{Queue}} \bibinfo{volume}{16},
  \bibinfo{number}{3} (\bibinfo{year}{2018}), \bibinfo{pages}{31--57}.
\newblock


\bibitem[Liu and Guo(2019)]%
        {liu2019bidirectional}
\bibfield{author}{\bibinfo{person}{Gang Liu} {and} \bibinfo{person}{Jiabao
  Guo}.} \bibinfo{year}{2019}\natexlab{}.
\newblock \showarticletitle{Bidirectional LSTM with attention mechanism and
  convolutional layer for text classification}.
\newblock \bibinfo{journal}{\emph{Neurocomputing}}  \bibinfo{volume}{337}
  (\bibinfo{year}{2019}), \bibinfo{pages}{325--338}.
\newblock


\bibitem[Liu et~al\mbox{.}(2018)]%
        {liu2018nlize}
\bibfield{author}{\bibinfo{person}{Shusen Liu}, \bibinfo{person}{Zhimin Li},
  \bibinfo{person}{Tao Li}, \bibinfo{person}{Vivek Srikumar},
  \bibinfo{person}{Valerio Pascucci}, {and} \bibinfo{person}{Peer-Timo
  Bremer}.} \bibinfo{year}{2018}\natexlab{}.
\newblock \showarticletitle{Nlize: A perturbation-driven visual interrogation
  tool for analyzing and interpreting natural language inference models}.
\newblock \bibinfo{journal}{\emph{IEEE transactions on visualization and
  computer graphics}} \bibinfo{volume}{25}, \bibinfo{number}{1}
  (\bibinfo{year}{2018}), \bibinfo{pages}{651--660}.
\newblock


\bibitem[Lundberg and Lee(2017)]%
        {lundberg2017unified}
\bibfield{author}{\bibinfo{person}{Scott~M Lundberg} {and}
  \bibinfo{person}{Su-In Lee}.} \bibinfo{year}{2017}\natexlab{}.
\newblock \showarticletitle{A Unified Approach to Interpreting Model
  Predictions}.
\newblock In \bibinfo{booktitle}{\emph{Advances in Neural Information
  Processing Systems 30}}, \bibfield{editor}{\bibinfo{person}{I.~Guyon},
  \bibinfo{person}{U.~V. Luxburg}, \bibinfo{person}{S.~Bengio},
  \bibinfo{person}{H.~Wallach}, \bibinfo{person}{R.~Fergus},
  \bibinfo{person}{S.~Vishwanathan}, {and} \bibinfo{person}{R.~Garnett}}
  (Eds.). \bibinfo{publisher}{Curran Associates, Inc.},
  \bibinfo{pages}{4765--4774}.
\newblock
\urldef\tempurl%
\url{http://papers.nips.cc/paper/7062-a-unified-approach-to-interpreting-model-predictions.pdf}
\showURL{%
\tempurl}


\bibitem[Ma et~al\mbox{.}(2017)]%
        {ma2017lamp}
\bibfield{author}{\bibinfo{person}{Shiqing Ma}, \bibinfo{person}{Yousra Aafer},
  \bibinfo{person}{Zhaogui Xu}, \bibinfo{person}{Wen-Chuan Lee},
  \bibinfo{person}{Juan Zhai}, \bibinfo{person}{Yingqi Liu}, {and}
  \bibinfo{person}{Xiangyu Zhang}.} \bibinfo{year}{2017}\natexlab{}.
\newblock \showarticletitle{LAMP: data provenance for graph based machine
  learning algorithms through derivative computation}. In
  \bibinfo{booktitle}{\emph{Proceedings of the 2017 11th Joint Meeting on
  Foundations of Software Engineering}}. \bibinfo{pages}{786--797}.
\newblock


\bibitem[Ma et~al\mbox{.}(2018)]%
        {ma2018mode}
\bibfield{author}{\bibinfo{person}{Shiqing Ma}, \bibinfo{person}{Yingqi Liu},
  \bibinfo{person}{Wen-Chuan Lee}, \bibinfo{person}{Xiangyu Zhang}, {and}
  \bibinfo{person}{Ananth Grama}.} \bibinfo{year}{2018}\natexlab{}.
\newblock \showarticletitle{MODE: automated neural network model debugging via
  state differential analysis and input selection}. In
  \bibinfo{booktitle}{\emph{Proceedings of the 2018 26th ACM Joint Meeting on
  European Software Engineering Conference and Symposium on the Foundations of
  Software Engineering}}. \bibinfo{pages}{175--186}.
\newblock


\bibitem[McLachlan and Basford(1988)]%
        {mclachlan1988mixture}
\bibfield{author}{\bibinfo{person}{Geoffrey~J McLachlan} {and}
  \bibinfo{person}{Kaye~E Basford}.} \bibinfo{year}{1988}\natexlab{}.
\newblock \bibinfo{booktitle}{\emph{Mixture models: Inference and applications
  to clustering}}. Vol.~\bibinfo{volume}{38}.
\newblock \bibinfo{publisher}{M. Dekker New York}.
\newblock


\bibitem[Ming et~al\mbox{.}(2017)]%
        {ming2017understanding}
\bibfield{author}{\bibinfo{person}{Yao Ming}, \bibinfo{person}{Shaozu Cao},
  \bibinfo{person}{Ruixiang Zhang}, \bibinfo{person}{Zhen Li},
  \bibinfo{person}{Yuanzhe Chen}, \bibinfo{person}{Yangqiu Song}, {and}
  \bibinfo{person}{Huamin Qu}.} \bibinfo{year}{2017}\natexlab{}.
\newblock \showarticletitle{Understanding hidden memories of recurrent neural
  networks}. In \bibinfo{booktitle}{\emph{2017 IEEE Conference on Visual
  Analytics Science and Technology (VAST)}}. IEEE, \bibinfo{pages}{13--24}.
\newblock


\bibitem[Molnar(2020)]%
        {molnar2020interpretable}
\bibfield{author}{\bibinfo{person}{Christoph Molnar}.}
  \bibinfo{year}{2020}\natexlab{}.
\newblock \bibinfo{booktitle}{\emph{Interpretable machine learning}}.
\newblock \bibinfo{publisher}{Lulu. com}.
\newblock


\bibitem[Montavon et~al\mbox{.}(2018)]%
        {montavon2018methods}
\bibfield{author}{\bibinfo{person}{Gr{\'e}goire Montavon},
  \bibinfo{person}{Wojciech Samek}, {and} \bibinfo{person}{Klaus-Robert
  M{\"u}ller}.} \bibinfo{year}{2018}\natexlab{}.
\newblock \showarticletitle{Methods for interpreting and understanding deep
  neural networks}.
\newblock \bibinfo{journal}{\emph{Digital Signal Processing}}
  \bibinfo{volume}{73} (\bibinfo{year}{2018}), \bibinfo{pages}{1--15}.
\newblock


\bibitem[Murugesan et~al\mbox{.}(2019)]%
        {murugesan2019deepcompare}
\bibfield{author}{\bibinfo{person}{Sugeerth Murugesan}, \bibinfo{person}{Sana
  Malik}, \bibinfo{person}{Fan Du}, \bibinfo{person}{Eunyee Koh}, {and}
  \bibinfo{person}{Tuan~Manh Lai}.} \bibinfo{year}{2019}\natexlab{}.
\newblock \showarticletitle{Deepcompare: Visual and interactive comparison of
  deep learning model performance}.
\newblock \bibinfo{journal}{\emph{IEEE computer graphics and applications}}
  \bibinfo{volume}{39}, \bibinfo{number}{5} (\bibinfo{year}{2019}),
  \bibinfo{pages}{47--59}.
\newblock


\bibitem[Okudono et~al\mbox{.}(2020)]%
        {okudono2020weighted}
\bibfield{author}{\bibinfo{person}{Takamasa Okudono}, \bibinfo{person}{Masaki
  Waga}, \bibinfo{person}{Taro Sekiyama}, {and} \bibinfo{person}{Ichiro
  Hasuo}.} \bibinfo{year}{2020}\natexlab{}.
\newblock \showarticletitle{Weighted automata extraction from recurrent neural
  networks via regression on state spaces}. In
  \bibinfo{booktitle}{\emph{Proceedings of the AAAI Conference on Artificial
  Intelligence}}, Vol.~\bibinfo{volume}{34}. \bibinfo{pages}{5306--5314}.
\newblock


\bibitem[Olah et~al\mbox{.}(2018)]%
        {olah2018building}
\bibfield{author}{\bibinfo{person}{Chris Olah}, \bibinfo{person}{Arvind
  Satyanarayan}, \bibinfo{person}{Ian Johnson}, \bibinfo{person}{Shan Carter},
  \bibinfo{person}{Ludwig Schubert}, \bibinfo{person}{Katherine Ye}, {and}
  \bibinfo{person}{Alexander Mordvintsev}.} \bibinfo{year}{2018}\natexlab{}.
\newblock \showarticletitle{The building blocks of interpretability}.
\newblock \bibinfo{journal}{\emph{Distill}} \bibinfo{volume}{3},
  \bibinfo{number}{3} (\bibinfo{year}{2018}), \bibinfo{pages}{e10}.
\newblock


\bibitem[Peng et~al\mbox{.}(2020)]%
        {peng2020first}
\bibfield{author}{\bibinfo{person}{Zi Peng}, \bibinfo{person}{Jinqiu Yang},
  \bibinfo{person}{Tse-Hsun Chen}, {and} \bibinfo{person}{Lei Ma}.}
  \bibinfo{year}{2020}\natexlab{}.
\newblock \showarticletitle{A first look at the integration of machine learning
  models in complex autonomous driving systems: A case study on apollo}. In
  \bibinfo{booktitle}{\emph{Proceedings of the 28th ACM Joint Meeting on
  European Software Engineering Conference and Symposium on the Foundations of
  Software Engineering}}. \bibinfo{pages}{1240--1250}.
\newblock


\bibitem[Quora(2019)]%
        {quora}
\bibfield{author}{\bibinfo{person}{Quora}.} \bibinfo{year}{2019}\natexlab{}.
\newblock \bibinfo{title}{Quora insincere questions classification}.
\newblock
\newblock
\urldef\tempurl%
\url{https://www.kaggle.com/c/quora-insincere-questions-classification/data}
\showURL{%
\tempurl}


\bibitem[Rabusseau et~al\mbox{.}(2019)]%
        {rabusseau2019connecting}
\bibfield{author}{\bibinfo{person}{Guillaume Rabusseau},
  \bibinfo{person}{Tianyu Li}, {and} \bibinfo{person}{Doina Precup}.}
  \bibinfo{year}{2019}\natexlab{}.
\newblock \showarticletitle{Connecting weighted automata and recurrent neural
  networks through spectral learning}. In \bibinfo{booktitle}{\emph{The 22nd
  International Conference on Artificial Intelligence and Statistics}}. PMLR,
  \bibinfo{pages}{1630--1639}.
\newblock


\bibitem[Rahnama and Bostr{\"o}m(2019)]%
        {rahnama2019study}
\bibfield{author}{\bibinfo{person}{Amir Hossein~Akhavan Rahnama} {and}
  \bibinfo{person}{Henrik Bostr{\"o}m}.} \bibinfo{year}{2019}\natexlab{}.
\newblock \showarticletitle{A study of data and label shift in the LIME
  framework}.
\newblock \bibinfo{journal}{\emph{arXiv preprint arXiv:1910.14421}}
  (\bibinfo{year}{2019}).
\newblock


\bibitem[Ribeiro et~al\mbox{.}(2016)]%
        {ribeiro2016should}
\bibfield{author}{\bibinfo{person}{Marco~Tulio Ribeiro},
  \bibinfo{person}{Sameer Singh}, {and} \bibinfo{person}{Carlos Guestrin}.}
  \bibinfo{year}{2016}\natexlab{}.
\newblock \showarticletitle{" Why should i trust you?" Explaining the
  predictions of any classifier}. In \bibinfo{booktitle}{\emph{Proceedings of
  the 22nd ACM SIGKDD international conference on knowledge discovery and data
  mining}}. \bibinfo{pages}{1135--1144}.
\newblock


\bibitem[Ribeiro et~al\mbox{.}(2018)]%
        {ribeiro2018anchors}
\bibfield{author}{\bibinfo{person}{Marco~Tulio Ribeiro},
  \bibinfo{person}{Sameer Singh}, {and} \bibinfo{person}{Carlos Guestrin}.}
  \bibinfo{year}{2018}\natexlab{}.
\newblock \showarticletitle{Anchors: High-precision model-agnostic
  explanations}. In \bibinfo{booktitle}{\emph{Proceedings of the AAAI
  conference on artificial intelligence}}, Vol.~\bibinfo{volume}{32}.
\newblock


\bibitem[Schneider et~al\mbox{.}(2021)]%
        {schneider2021explain}
\bibfield{author}{\bibinfo{person}{Tobias Schneider}, \bibinfo{person}{Joana
  Hois}, \bibinfo{person}{Alischa Rosenstein}, \bibinfo{person}{Sabiha
  Ghellal}, \bibinfo{person}{Dimitra Theofanou-F{\"u}lbier}, {and}
  \bibinfo{person}{Ansgar~RS Gerlicher}.} \bibinfo{year}{2021}\natexlab{}.
\newblock \showarticletitle{ExplAIn Yourself! Transparency for Positive UX in
  Autonomous Driving}. In \bibinfo{booktitle}{\emph{Proceedings of the 2021 CHI
  Conference on Human Factors in Computing Systems}}. \bibinfo{pages}{1--12}.
\newblock


\bibitem[Selvaraju et~al\mbox{.}(2017)]%
        {selvaraju2017grad}
\bibfield{author}{\bibinfo{person}{Ramprasaath~R Selvaraju},
  \bibinfo{person}{Michael Cogswell}, \bibinfo{person}{Abhishek Das},
  \bibinfo{person}{Ramakrishna Vedantam}, \bibinfo{person}{Devi Pariah}, {and}
  \bibinfo{person}{Dhruv Batra}.} \bibinfo{year}{2017}\natexlab{}.
\newblock \showarticletitle{Grad-cam: Visual explanations from deep networks
  via gradient-based localization}. In \bibinfo{booktitle}{\emph{Proceedings of
  the IEEE international conference on computer vision}}.
  \bibinfo{pages}{618--626}.
\newblock


\bibitem[Simonyan et~al\mbox{.}(2013)]%
        {simonyan2013deep}
\bibfield{author}{\bibinfo{person}{Karen Simonyan}, \bibinfo{person}{Andrea
  Vedaldi}, {and} \bibinfo{person}{Andrew Zisserman}.}
  \bibinfo{year}{2013}\natexlab{}.
\newblock \showarticletitle{Deep inside convolutional networks: Visualising
  image classification models and saliency maps}.
\newblock \bibinfo{journal}{\emph{arXiv preprint arXiv:1312.6034}}
  (\bibinfo{year}{2013}).
\newblock


\bibitem[Smilkov et~al\mbox{.}(2017)]%
        {smilkov2017smoothgrad}
\bibfield{author}{\bibinfo{person}{Daniel Smilkov}, \bibinfo{person}{Nikhil
  Thorat}, \bibinfo{person}{Been Kim}, \bibinfo{person}{Fernanda Vi{\'e}gas},
  {and} \bibinfo{person}{Martin Wattenberg}.} \bibinfo{year}{2017}\natexlab{}.
\newblock \showarticletitle{Smoothgrad: removing noise by adding noise}.
\newblock \bibinfo{journal}{\emph{arXiv preprint arXiv:1706.03825}}
  (\bibinfo{year}{2017}).
\newblock


\bibitem[Strobelt et~al\mbox{.}(2017)]%
        {strobelt2017lstmvis}
\bibfield{author}{\bibinfo{person}{Hendrik Strobelt},
  \bibinfo{person}{Sebastian Gehrmann}, \bibinfo{person}{Hanspeter Pfister},
  {and} \bibinfo{person}{Alexander~M Rush}.} \bibinfo{year}{2017}\natexlab{}.
\newblock \showarticletitle{Lstmvis: A tool for visual analysis of hidden state
  dynamics in recurrent neural networks}.
\newblock \bibinfo{journal}{\emph{IEEE transactions on visualization and
  computer graphics}} \bibinfo{volume}{24}, \bibinfo{number}{1}
  (\bibinfo{year}{2017}), \bibinfo{pages}{667--676}.
\newblock


\bibitem[Vaswani et~al\mbox{.}(2017)]%
        {vaswani2017attention}
\bibfield{author}{\bibinfo{person}{Ashish Vaswani}, \bibinfo{person}{Noam
  Shazeer}, \bibinfo{person}{Niki Parmar}, \bibinfo{person}{Jakob Uszkoreit},
  \bibinfo{person}{Llion Jones}, \bibinfo{person}{Aidan~N Gomez},
  \bibinfo{person}{{\L}ukasz Kaiser}, {and} \bibinfo{person}{Illia
  Polosukhin}.} \bibinfo{year}{2017}\natexlab{}.
\newblock \showarticletitle{Attention is all you need}.
\newblock \bibinfo{journal}{\emph{Advances in neural information processing
  systems}}  \bibinfo{volume}{30} (\bibinfo{year}{2017}).
\newblock


\bibitem[Weiss et~al\mbox{.}(2018)]%
        {weiss2018extracting}
\bibfield{author}{\bibinfo{person}{Gail Weiss}, \bibinfo{person}{Yoav
  Goldberg}, {and} \bibinfo{person}{Eran Yahav}.}
  \bibinfo{year}{2018}\natexlab{}.
\newblock \showarticletitle{Extracting automata from recurrent neural networks
  using queries and counterexamples}. In
  \bibinfo{booktitle}{\emph{International Conference on Machine Learning}}.
  PMLR, \bibinfo{pages}{5247--5256}.
\newblock


\bibitem[Xie et~al\mbox{.}(2021)]%
        {xie2021rnnrepair}
\bibfield{author}{\bibinfo{person}{Xiaofei Xie}, \bibinfo{person}{Wenbo Guo},
  \bibinfo{person}{Lei Ma}, \bibinfo{person}{Wei Le}, \bibinfo{person}{Jian
  Wang}, \bibinfo{person}{Lingjun Zhou}, \bibinfo{person}{Yang Liu}, {and}
  \bibinfo{person}{Xinyu Xing}.} \bibinfo{year}{2021}\natexlab{}.
\newblock \showarticletitle{RNNrepair: Automatic RNN repair via model-based
  analysis}. In \bibinfo{booktitle}{\emph{International Conference on Machine
  Learning}}. PMLR, \bibinfo{pages}{11383--11392}.
\newblock


\bibitem[Yu et~al\mbox{.}(2022)]%
        {yu2020deeprepair}
\bibfield{author}{\bibinfo{person}{Bing Yu}, \bibinfo{person}{Hua Qi},
  \bibinfo{person}{Qing Guo}, \bibinfo{person}{Felix Juefei-Xu},
  \bibinfo{person}{Xiaofei Xie}, \bibinfo{person}{Lei Ma}, {and}
  \bibinfo{person}{Jianjun Zhao}.} \bibinfo{year}{2022}\natexlab{}.
\newblock \showarticletitle{DeepRepair: Style-Guided Repairing for Deep Neural
  Networks in the Real-World Operational Environment}.
\newblock \bibinfo{journal}{\emph{IEEE Transactions on Reliability}}
  \bibinfo{volume}{71}, \bibinfo{number}{4} (\bibinfo{year}{2022}),
  \bibinfo{pages}{1401--1416}.
\newblock
\urldef\tempurl%
\url{https://doi.org/10.1109/TR.2021.3096332}
\showDOI{\tempurl}


\bibitem[Zeiler and Fergus(2014)]%
        {zeiler2014visualizing}
\bibfield{author}{\bibinfo{person}{Matthew~D Zeiler} {and} \bibinfo{person}{Rob
  Fergus}.} \bibinfo{year}{2014}\natexlab{}.
\newblock \showarticletitle{Visualizing and understanding convolutional
  networks}. In \bibinfo{booktitle}{\emph{European conference on computer
  vision}}. Springer, \bibinfo{pages}{818--833}.
\newblock


\bibitem[Zhang et~al\mbox{.}(2015)]%
        {zhang2015character}
\bibfield{author}{\bibinfo{person}{Xiang Zhang}, \bibinfo{person}{Junbo Zhao},
  {and} \bibinfo{person}{Yann LeCun}.} \bibinfo{year}{2015}\natexlab{}.
\newblock \showarticletitle{Character-level convolutional networks for text
  classification}.
\newblock \bibinfo{journal}{\emph{Advances in neural information processing
  systems}}  \bibinfo{volume}{28} (\bibinfo{year}{2015}).
\newblock


\end{thebibliography}
